\documentclass[lettersize,journal]{IEEEtran}
\usepackage{amsmath,amsfonts}
\usepackage{algorithmic}
\usepackage{algorithm}
\usepackage{array}
\usepackage[caption=false,font=normalsize]{subfig}
\usepackage{textcomp}
\usepackage{stfloats}
\usepackage{url}
\usepackage{verbatim}
\usepackage{graphicx}
\usepackage{cite}
\usepackage{xcolor}
\newtheorem{remark}{Remark}

\begin{document}

\title{Design of 3D Beamforming and Deployment Strategies for ISAC-Based HAPS Systems}

\author{
Xue Zhang,  {\em Student Member, IEEE}, Bang Huang,  {\em Member, IEEE}, \\ and Mohamed-Slim Alouini, {\em Fellow, IEEE}
\thanks{The authors are with Computer, Electrical and Mathematical Sciences
and Engineering (CEMSE) Division, Department of Electrical and Computer
Engineering, King Abdullah University of Science and Technology (KAUST), Thuwal 23955-6900, Saudi Arabia. (e-mail: xue.zhang@kaust.edu.sa; bang.huang@kaust.edu.sa; slim.alouini@kaust.edu.sa) (Corresponding author: Bang Huang). 
}
\vspace{-8mm}
}


\maketitle

\begin{abstract}
This paper investigates high-altitude platform station (HAPS) systems empowered by integrated sensing and communication (ISAC), where a HAPS simultaneously delivers communication services to multiple users while conducting synthetic aperture radar (SAR) imaging for ground target sensing. Considering the operational characteristics of SAR, we examine two deployment strategies: (i) a quasi-stationary HAPS that hovers at an optimized location during SAR operation, following the stop-and-go scanning model, and (ii) a dynamic HAPS that continuously adapts its trajectory along a circular path. For both strategies, our objective is to maximize the weighted sum-rate of communication users while ensuring SAR imaging performance requirements, including beampattern gain and signal-to-noise ratio (SNR), are satisfied. To this end, we jointly optimize the HAPS deployment strategy, either three-dimensional (3D) placement or trajectory, together with transmit beamforming, subject to practical constraints such as transmit power, energy consumption, and flight dynamics. The resulting optimization problems are inherently non-convex. To overcome this challenge, we develop efficient algorithms that combine convex and non-convex optimization techniques to obtain high-quality suboptimal solutions. Numerical results validate the effectiveness of the proposed approaches and demonstrate their performance advantages over benchmark schemes.
\end{abstract}


\section{Introduction}
\label{sec_introdution}
High-altitude platform station (HAPS) has gained significant attention in recent years owing to their wide coverage area, reliable line-of-sight communication links, and fixed positions relative to Earth~\cite{abbasi2024haps,yahia2022haps}. Flying at high-altitude stratospheric levels between 20 and 50 kilometers, HAPS serves as an essential bridge between terrestrial and satellite communication systems~\cite{kurt2021vision}. These platforms are highly versatile, supporting various applications such as wide-area surveillance, environmental monitoring, and communication services~\cite{mohammed2011role}.

The rapid advancement of next-generation wireless networks, coupled with the explosive growth in wireless devices, has led to an unprecedented demand for wireless spectrum. At the same time, integrating sensing capabilities into communication networks introduces additional resource requirements, including spectrum, energy, and hardware. To address these challenges, the concept of integrated sensing and communication (ISAC) has emerged, combining sensing and communication within a unified framework to enhance efficiency~\cite{wang2024network,zhang2024joint,liu2022integrated,benaya2025aerial,kaushik2024integrated,zhang2025ris}. Radio frequency (RF) imaging serves as a fundamental enabler for ISAC, offering high-resolution sensing capabilities that remain effective across a wide range of conditions, including adverse weather and continuous day–night operation. Incorporating RF imaging into sixth-generation (6G) networks facilitates simultaneous communication and imaging services, which are essential for emerging applications such as low-altitude space economy scenarios~\cite{zheng2024random}. 

HAPS are particularly well-suited to support ISAC due to their unique characteristics. In particular, synthetic aperture radar (SAR), a well-established RF imaging technology, can be deployed on HAPSs to deliver high-resolution imaging from the stratosphere~\cite{wang2014high,huang2025design}. This paper focuses on developing a joint SAR imaging and communication framework for HAPS. By leveraging the advantages of HAPS and the principles of ISAC, the proposed framework aims to address the growing demand for simultaneous communication and sensing in next-generation networks.

\subsection{Prior Work}
The implementation of HAPS faces a series of technical challenges, particularly in the domains of aerodynamics and communication system design. In aerodynamic design, the HAPS system must exhibit several essential characteristics, including a lightweight and robust structure, resistance to harsh environmental conditions, adequate payload capacity, high-precision positioning, stable station-keeping flight, and reliable autonomous navigation to support long-endurance missions~\cite{gsma2021high}. Additionally, many sensing, monitoring, and communication applications demand higher payload capacity to support various sensors and base stations. At the same time, the communication system must include accurate link budgeting, efficient multiple access techniques, and effective interference management to ensure stable, reliable service over a wide area. 

To overcome these limitations, prior research on HAPS platforms has proposed a range of approaches, including trajectory optimization, efficient resource scheduling, system performance evaluation, and comprehensive link and power budget planning~\cite{nauman2017system,marriott2020trajectory,azzahra2019noma,ji2020energy,hsieh2020uav,javed2023interdisciplinary}. For instance, aiming at minimizing both error rates and power consumption, the authors in~\cite{nauman2017system} carried out link budgeting with quadrature phase shift keying (QPSK) modulation with varying code rates. The flight trajectory was proposed in~\cite{marriott2020trajectory} for a solar-powered, high-altitude, and long-endurance aircraft. The work of~\cite{azzahra2019noma,ji2020energy} investigated the non-orthogonal multiple access (NOMA) technology for more efficient resource management. In~\cite{hsieh2020uav}, a repetitive flight pattern combined with a steerable adaptive antenna array was proposed to ensure uninterrupted service to users located within the same coverage cell. To resolve the interdependent issues of flight dynamics and communication performance in HAPS networks,~\cite{javed2023interdisciplinary} developed a holistic approach that integrates knowledge from atmospheric science, aerodynamics, wireless communications, and renewable energy. This comprehensive strategy supports optimal operational planning and long-endurance performance through effective resource allocation.

Based on the aforementioned discussion, we explored the key challenges and potential solutions in aerodynamics and communication for HAPS systems. As HAPS platforms are increasingly recognized for their ability to provide wide-area coverage and efficient communication, a key challenge lies in effectively integrating communication and remote sensing imaging~\cite{wang2014mimo,wang2019first}. In particular, the implementation of integration of communication and SAR imaging (JCASAR) on HAPS systems is essential. This integration not only maximizes the multifunctionality of HAPS but also allows for optimal resource allocation across various applications, including communication, remote sensing, and surveillance. 

During the past few years, significant research has focused on JCASAR systems. For instance,~\cite{wang2019first} proposed an airborne multiple-input multiple-output (MIMO) radar capable of simultaneous data transmission and high-resolution SAR imaging without intramodal interference. In~\cite{tan2022joint}, a time-frequency spectrum shaping method was introduced to enable efficient waveform design for JCASAR. Furthermore,~\cite{yang2022waveform} presented a comprehensive watermarking framework tailored for such systems. Most of the existing methods implement JCASAR by employing orthogonal resource allocation, where communication and SAR imaging functionalities are separated in either the frequency or time domain. Although this strategy helps avoid interference, it often results in suboptimal resource utilization compared to fully integrated waveform designs~\cite{liu2022integrated}. Achieving JCASAR with unified waveforms introduces two major challenges. First, the waveform must exhibit sufficient randomness to support high-rate data transmission, which may negatively impact SAR imaging quality. Second, it is critical to strike an effective balance between communication performance and imaging resolution. To address these challenges,~\cite{zheng2024random} proposed a novel waveform design approach that explicitly incorporates data randomness while maintaining the dual functionality of communication and sensing, thereby enhancing the overall performance of JCASAR systems.

\subsection{Main Contributions}
To the best of our knowledge, our paper is the first to jointly investigate three-dimensional (3D) transmit beamforming design and deployment strategies for HAPS systems performing simultaneous SAR imaging and multi-user communication. The main contributions are summarized as follows:
\begin{itemize}
    \item \textbf{System Modeling Framework:} Unlike existing ISAC-enabled HAPS models, our framework explicitly integrates SAR imaging performance metrics with aerodynamic constraints, thereby ensuring reliable imaging quality while satisfying energy consumption requirements. 
    \item \textbf{Dual Deployment Strategies:} Leveraging the operational characteristics of SAR, we propose two deployment strategies for ISAC-enabled HAPS systems: (i) a \emph{quasi-stationary strategy}, where the HAPS remains fixed at an optimized location during SAR imaging, following the stop-and-go scanning model; and (ii) a \emph{dynamic strategy}, where the HAPS  equipped with circular SAR follows a circular flight trajectory.   
    \item \textbf{Quasi-Stationary HAPS Optimization:} For the quasi-stationary case, we formulate a joint optimization problem that simultaneously determines the HAPS placement and 3D transmit beamforming design, with the goal of maximizing the weighted sum-rate throughput of communication users while satisfying SAR imaging constraints and transmit power limitations. Due to the inherent non-convexity caused by the strong coupling between placement and beamforming, we propose an efficient algorithmic framework combining 3D grid search, successive convex approximation (SCA), and semidefinite relaxation (SDR), yielding high-quality suboptimal solutions.  
    \item \textbf{Dynamic HAPS Optimization:} For the dynamic scenario, we formulate a joint optimization problem involving both the HAPS trajectory and 3D transmit beamforming, aiming to maximize the average weighted sum-rate throughput while ensuring SAR imaging requirements, trajectory feasibility, energy consumption, and transmit power constraints. The problem is highly challenging due to the nonlinear dependence of the trajectory on steering vectors. To address this, we develop an alternating optimization framework that incorporates a trust-region-based SCA method, where Taylor-series linearization is applied to handle non-concave objectives and non-convex constraints, while trust regions guarantee approximation accuracy.  
    \item \textbf{Performance Validation:} We conduct numerical experiments to verify the effectiveness of the proposed methods and to highlight their advantages for deployment design over benchmark approaches that consider either communication-only or sensing-only designs.
\end{itemize}

\subsection{Outline and Notations}
\textit{Outline}: The remainder of the paper is organized as follows. Section \ref{sec_model} introduces the ISAC-enabled HAPS system model. Section~\ref{sec_formulation} constructs two optimization problems aimed at maximizing communication performance under SAR imaging constraints, each corresponding to a distinct HAPS deployment strategy. Sections~\ref{sec_quasi} and~\ref{sec_mobile} propose efficient solution methods for the respective problems. Numerical results about the proposed methods are provided in Section~\ref{sec_sim}, and Section~\ref{sec_con} finally concludes the paper.

\textit{Notations}: The vectors (matrices) are denoted by lower-case (upper-case) boldface characters. $\mathbf{I}_{M}$ stands for the $M\times M$ identity matrix, and $\mathbf{0}$ represents a zero matrix with appropriate dimension. $\dot{\mathbf{x}}(t)$ denotes the derivative of the time-dependent function $\mathbf{x}(t)$ with respect to the time $t$. Superscripts $T$, and $H$ represent the transpose and conjugate transpose, respectively. $\mathbb{E}\{\cdot\}$ represents the expectation operator. The trace of $\mathbf{A}$ are represented by $\mathrm{tr}(\mathbf{A})$. $[\mathbf{A}]_{i,j}$ stands for the $(i,j)$ entry of $\mathbf{A}$. $\mathbf{A}\succeq\mathbf{0}$ is equivalent to $\mathbf{A}$ being positive semidefinite. $|\cdot|$ stand for the magnitude of a complex number, and $\|\cdot\|$ is the Euclidean norm of a complex vector. $\mathcal{O}(\cdot)$ denotes the convolution Big-O notation.

\section{System Model}
\label{sec_model}
\begin{figure}[t]
\centering
\includegraphics[width=0.9\linewidth]{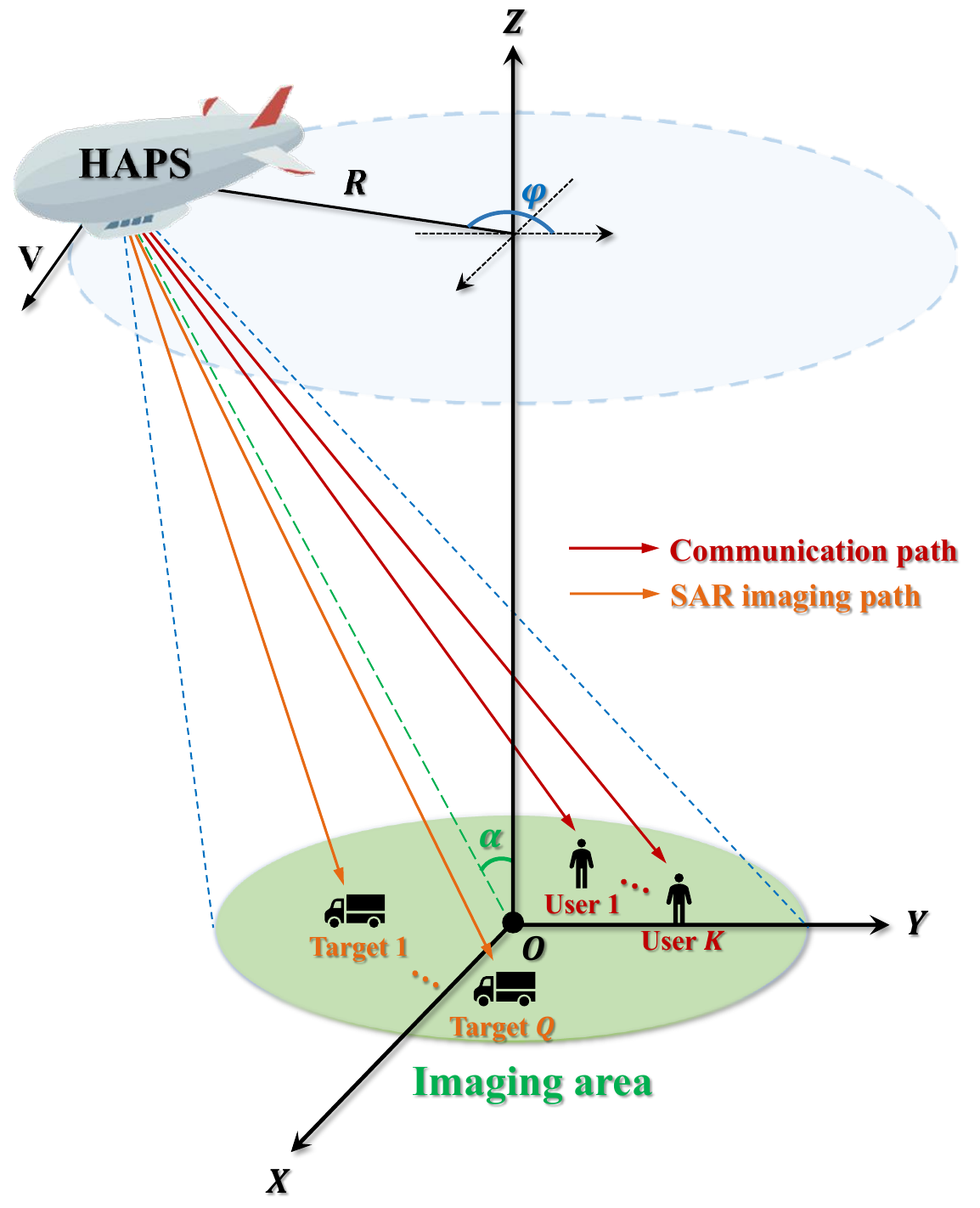}
\caption{ISAC-enabled HAPS system model.}
\label{system_model}
\end{figure}

As illustrated in Fig.~\ref{system_model}, we consider a HAPS system equipped with ISAC capability, in which the HAPS is equipped with a circular SAR system~\cite{soumekh1996reconnaissance,vachon2005digital} to simultaneously provide downlink communication for $K\geq 1$ users and high-resolution imaging for $Q\geq 1$ targets within the ground imaging region. The system adopts a 3D Cartesian coordinate framework, where the HAPS moves along a circular synthetic aperture with radius $R$ and tangential velocity $V$ in the $XOY$ plane, centered around the $Z$-axis. This circular flight trajectory enables beam-focused SAR imaging. During each operation period of duration $T > 0$ seconds, the SAR beam remains continuously directed toward the system center, denoted by point $O$. Each ground user is equipped with a single receive antenna, and the HAPS carries a vertically oriented uniform linear array (ULA) comprising $M$ antennas spaced at half-wavelength intervals, i.e., $d=\lambda/2$, with $\lambda$ denoting the carrier wavelength. 

For $k$th ground user with $k=1,2,\ldots,K$, its horizontal coordinate is fixed and represented by $\mathbf{m}_k=[x_k,y_k]^T$. Similarly, for $q$th ground target with $q=1,2,\ldots,Q$, its horizontal coordinate is fixed and denoted by $\mathbf{t}_q=[x_q,y_q]^T$. The HAPS is assumed to fly at a time-varying altitude $z(t)$, with a corresponding horizontal position at time $t$ given by $\mathbf{h}(t)=[x(t),y(t)]^T$, where $0\leq t\leq T$. At the initial time $t=0$, the HAPS is positioned on the positive half of the $Y$-axis. At the end of each period $T$, it returns to this initial location to enable periodic SAR imaging and ensure continuous and consistent communication in subsequent cycles. Thus, the trajectories of the HAPS need to satisfy $\mathbf{h}(0) = \mathbf{h}(T)$ and $z[0]=z[T]$. Meanwhile, the azimuth angle $\varphi(t)$ is defined relative to the $Y$ axis as $\varphi(t)=\frac{Vt}{R}$. In practice applications, the movement of HAPS is constrained by the maximum speed constraints, i.e., $\|\dot{\mathbf{h}}(t)\|=V_{xy}\leq V_{xy\mathrm{max}}$ and $\|\dot{z}(t)\|=V_z\leq V_{z\mathrm{max}}$. Here, $V_{xy}$ and $V_z$ represent the horizontal and vertical speed of the HAPS in meters per second (m/s), respectively, and $V_{xy\mathrm{max}}$ and $V_{z\mathrm{max}}$ are their respective maximum values. 

To facilitate analysis and optimization, the period $T$ is divided into $N$ equal time slots, indexed by $n = 0,1,\ldots, N$. Each time slot has a duration of $\Delta_t = \frac{T}{N}$, which is chosen to be sufficiently small such that the HAPS position can be considered approximately constant within each slot, even when operating at its maximum speed. Let $\alpha\in(0,\pi/2)$ denote the fixed observation angle of the SAR onboard the HAPS. Based on this configuration, the HAPS trajectory in the horizontal plane can be expressed as $x[n]=R[n]\sin(\varphi[n])$, and $y[n]=R[n]\cos(\varphi[n])$, where $R[n]=z[n]\tan(\alpha)$, and $\varphi[n]=\frac{Vn\Delta_t}{R}$~\cite{hu2022trajectory}. As suggested in~\cite{wu2018joint}, the number of time slots $N$ should be carefully selected to balance the trade-off between accuracy and computational complexity. To ensure that the discretized trajectory satisfies a predefined accuracy threshold $\varepsilon$, the minimum required value of $N$ is given by $N\ge\frac{V_{xy}T}{H_{\mathrm{min}}\varepsilon}$, where $H_{\mathrm{min}}$ is the minimum height of the HAPS. Moreover, the aforementioned trajectory constraints can be equivalently written as~\cite{hua20203d,zeng2017energy}
\begin{align}
    &\mathbf{h}[0] = \mathbf{h}[N],\,\, z[0] = z[N],\,\,H_{\mathrm{min}} \leq z[n] \leq H_{\mathrm{max}}, \\
    &\|\mathbf{h}[n+1]-\mathbf{h}[n]\|=V_{xy} \leq V_{xy\mathrm{max}}\Delta_t, \\
    &\|z[n+1]-z[n]\|= V_z\Delta_t \leq V_{z\mathrm{max}}\Delta_t,
\end{align}
where $H_{\mathrm{max}}$ is the maximum height of the HAPS, $V_{xy}\Delta_t$ and $V_z\Delta_t$ are the maximum horizontal and vertical distance that the HAPS can travel in each time slot, respectively. 

Considering that the HAPS simultaneously transmits dedicated SAR imaging signals and communication signals intended for ground users, the transmitted signal at time slot $n$ is given by  
\begin{align}\label{transmitted signal}
    \mathbf{x}(n) = \sum_{k=1}^K\mathbf{w}_k[n]s_k[n] + \mathbf{s}_{0}[n],  
\end{align}
where $\mathbf{s}_0[n]\in\mathbb{C}^{M\times 1}$ denotes the dedicated SAR imaging signal at time slot $n$, assumed to be a zero-mean independent random vector with covariance matrix $\pmb{\mathcal{R}}_s[n]=\mathbb{E}\{\mathbf{s}_0[n]\mathbf{s}_0^H[n]\}\succeq \pmb{0}$, $s_k[n]\sim\mathcal{CN}(0,1)$ represents the desired communication signal intended by the $k$th user, described as zero-mean, unit-variance circularly symmetric complex Gaussian random variable, and $\mathbf{w}_k[n]\in\mathbb{C}^{M\times 1}$ is its corresponding transmit beamforming vector. Therefore, the average transmit power at slot $n$ is obtained as 
\begin{align}\label{average transmit power}
    P_{\mathrm{ave}}[n] =& \mathbb{E}\{\|\mathbf{x}[n]\|^2\} = \sum_{k=1}^K\|\mathbf{w}_k[n]\|^2 + \mathrm{tr}(\pmb{\mathcal{R}}_s[n]). 
\end{align}
Suppose that the maximum allowable transmit power at the HAPS is $P_{\mathrm{max}}$, the corresponding power constraint can be given by $P_{\mathrm{ave}}[n]\leq P_{\mathrm{max}}$.

Due to the high altitude of the HAPS, a strong line-of-sight (LOS) link typically exists between the platform and each ground user. However, in obstructed environments, scattering and reflection effects may also occur. To accurately capture these characteristics, the channel is modeled as a 3D Rician fading channel, consisting of a dominant LOS component alongside additional non-LOS (NLOS) components. Accordingly, the channel vector between the HAPS and the $k$th user at time slot $n$ is written as~\cite{shamsabadi2024enhancing,abbasi2024hemispherical} 
\begin{align}
    &\mathbf{g}(\mathbf{h}[n],z[n],\mathbf{m}_k) \notag\\
    &\quad=\frac{1}{\sqrt{\ell}}\left(\sqrt{\frac{K_u}{K_u+1}}\mathbf{g}_{k,\mathrm{LOS}}(n)+\sqrt{\frac{1}{K_u+1}}\mathbf{g}_{k,\mathrm{NLOS}}(n)\right),
\end{align}
where $K_u$ is the Rician factor, $\ell=d^2(\mathbf{h}[n],z[n],\mathbf{m}_k)/\rho_0$ denotes the free-space path loss with $\rho_0=(\frac{4\pi}{\lambda})^2$ and $d(\mathbf{h}[n],z[n],\mathbf{m}_k)=\sqrt{\|\mathbf{h}[n]-\mathbf{m}_k\|^2+z[n]^2}$ being the distance between the HAPS and the $k$th user at time slot $n$, $\mathbf{g}_{k,\mathrm{NLOS}}(n)$ represents the NLOS component, whose elements are independently drawn from a complex Gaussian distribution with zero mean and unit variance, and $\mathbf{g}_{k,\mathrm{LOS}}(n)=\mathbf{a}(\mathbf{h}[n],z[n],\mathbf{m}_k)$ denotes the LOS component, represented by the steering vector corresponding to the $k$th user, i.e.,
\begin{align}\label{steering vector}
    \mathbf{a}(\mathbf{h}[n],z[n],\mathbf{m}_k) =& \left[1,e^{j\pi\cos(\theta(\mathbf{h}[n],z[n],\mathbf{m}_k)}),\right.\notag\\
    &\quad\left.\ldots,e^{j\pi(M-1)\cos(\theta(\mathbf{h}[n],z[n],\mathbf{m}_k))}\right]^T. 
\end{align}
Here, $\theta(\mathbf{h}[n],z[n],\mathbf{m}_k)$ is the angle of departure (AoA) towards the $k$th user, given by 
\begin{align}
    \theta(\mathbf{h}[n],z[n],\mathbf{m}_k) = \arccos\frac{z[n]}{\sqrt{\|\mathbf{h}[n]-\mathbf{m}_k\|^2+z[n]^2}}.
\end{align}
Consequently, the received signal at the $k$th user during time slot $n$ can be written as
\begin{align}\label{received signal}
    y_k[n] = \mathbf{g}_k^H(\mathbf{h}[n],z[n],\mathbf{m}_k)\mathbf{x}[n] + n_k[n],
\end{align}
where $n_k[n]\sim\mathcal{CN}(0,\sigma_k^2)$ denotes the additive white Gaussian noise, with $\sigma_k^2$ being the noise power.

\begin{figure*}[t!]
\centering
\subfloat[SHF]{
\includegraphics[width=0.34\linewidth]{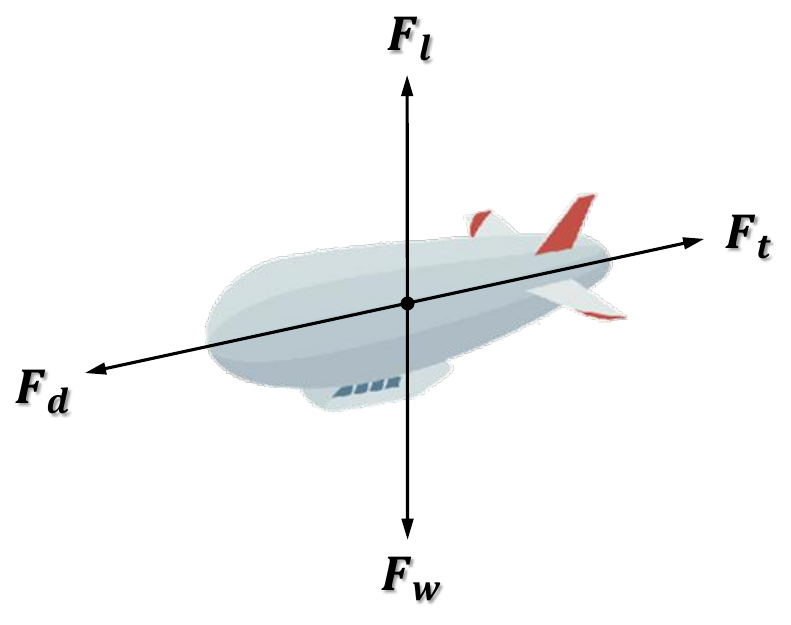}}
\quad
\subfloat[SCF at a bank angle $\vartheta$]{
\includegraphics[width=0.5050\linewidth]{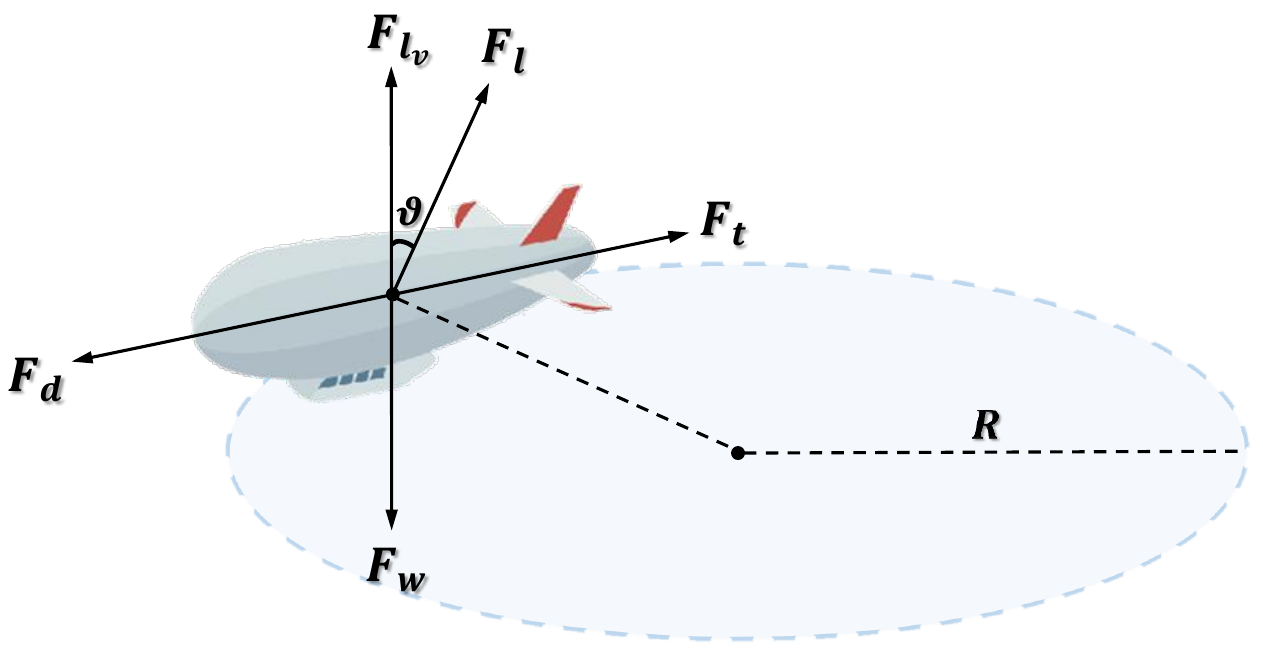}}
\caption{Forces acting on the HAPS.}
\label{SHCF}
\end{figure*}

\subsection{Aerodynamics}
From an aerodynamics perspective, maintaining steady circular flight (SCF) for station-keeping at a fixed altitude requires the HAPS to generate adequate propulsion power. As illustrated in Fig.\ref{SHCF}, SCF is primarily achieved through steady horizontal flight (SHF) with a constant bank angle $\vartheta$, where the vertical lift component $F_{lv}$ balances the aircraft’s weight $F_w$. During SHF, the lift force $F_l$ counteracts the gravitational force $F_w$, while the thrust $F_t$ compensates for the drag force $F_d$. Inspired by\cite{stengel2004flight,arum2020energy,javed2023interdisciplinary}, the propulsion power required for SHF at time slot $n$, denoted by $P_{\mathrm{SHF}}[n]$, depends on the true airspeed of the HAPS, $V = \sqrt{V_{xy}^2 + V_z^2}$, and the thrust $F_t$, expressed as
\begin{align}
    P_{\mathrm{SHF}}[n] = \frac{F_t[n]V}{\eta_p\eta_e},
\end{align}
where $\eta_p$ and $\eta_e$ are the propeller and engine efficiencies, respectively, and the thrust $F_t[n]$ at time slot $n$ is given by 
\begin{align}
    F_t[n] = \frac{1}{2}\rho_h[n]V^2SC_{D_0} + \frac{2\epsilon F_w^2}{\rho_h[n]SV^2}. 
\end{align}
Here, $\rho_h[n]$ denotes the air density at time slot $n$, approximated for the relevant altitude range by the curve-fitting expression $(0.95162z[n]^2-52.29356z[n]+753.39927)\times 10^{-3}$\,$kg/m^3$, $S$ is wing surface area, $C_{D_0}$ is the zero-lift drag coefficient, and $\epsilon=(\pi e_oAR_w)^{-1}$ with Oswald's efficiency factor $e_o$ and wing aspect ratio $AR_w$. The propulsion power $P_{\mathrm{SCF}}[n]$ is then calculated as
\begin{align}\label{P_SCF}
    P_{\mathrm{SCF}}[n] = \left(\frac{1}{\cos(\vartheta)}\right)^2P_{\mathrm{SHF}}[n]. 
\end{align}
The gradient of $P_{\mathrm{SCF}}[n]$ with respect to HAPS speed $V$ is written as
\begin{align}
    \frac{\partial P_{\mathrm{SCF}}[n]}{\partial V} = \frac{\Delta t}{\cos(\vartheta)^2f_pf_e}\left(\frac{3}{2}\rho_h[n]V^2SC_{D_0}-\frac{2\epsilon F_{\omega}^2}{\rho_h[n]SV^2}\right),
\end{align}
according to~\cite{javed2023interdisciplinary}, $P_{\mathrm{SCF}}[n]$ is strictly increasing in $V$. To simplify the energy constraint in Section~\ref{sec_model}-D, we observe that $\frac{1}{2}\rho_h[n]V^2SC_{D_0} \gg  \frac{2\epsilon F_w^2}{\rho_h[n]SV^2}$. Therefore, we can approximate $F_t[n]$ as $F_t[n]\approx\frac{1}{2}\rho_h[n]V^2SC_{D_0}$.

\begin{figure*}[ht!]
\begin{align}\label{gamma_k}
    \gamma_k(\mathbf{h}[n],z[n],\{\mathbf{w}_k[n]\},\pmb{\mathcal{R}}_s[n]) = \frac{\mathbb{E}\left\{\left|\mathbf{g}_k^H(\mathbf{h}[n],z[n],\mathbf{m}_k)\mathbf{w}_k[n]\right|^2\right\}}{\mathbb{E}\left\{\sum\limits_{p=1\atop p\neq k}^K\left|\mathbf{g}_k^H(\mathbf{h}[n],z[n],\mathbf{m}_k)\mathbf{w}_p[n]\right|^2\right\}+\mathbb{E}\left\{|\mathbf{g}_k^H(\mathbf{h}[n],z[n],\mathbf{m}_k)\mathbf{s}_0[n]|^2\right\}+\sigma_k^2}.
\end{align}
\hrulefill
\vspace*{10pt}
\end{figure*}

\subsection{Communication Metric}
From (\ref{received signal}), it is clear that each ground user suffers from co-channel interference caused by the communication signals $s_i[n]$ ($p\neq k$) intended for other users, as well as from the dedicated sensing signal $\mathbf{s}_0[n]$. Then the received signal-to-interference-plus-noise ratio (SINR) at the $k$th user during time slot $n$ can be expressed as (\ref{gamma_k}). Under a Gaussian noise environment, the achievable rate for the $k$th user at slot $n$, measured in bits-per-second-per-Hertz (bps/Hz), is given by 
\begin{align}
    R_k[n] = \log_2(1+\gamma_k(\mathbf{h}[n],z[n],\{\mathbf{w}_k[n]\},\pmb{\mathcal{R}}_s[n])). 
\end{align}

\subsection{SAR Imaging Metric}
With a finite set of $Q$ points, if the existence and exact positions of targets are unknown, these points are uniformly sampled across the entire region of interest. Alternatively, when the HAPS employs the SAR-GMTI technique~\cite{li2011influence,makhoul2014performance} for target tracking and approximate target positions are available, the points $\mathbf{t}_q$ can be set to these estimated locations. The SAR imaging process exploits both the sensing signal $\mathbf{s}_0$ and communication signals $s_k$ for $k=1, \ldots, K$. The imaging performance is then characterized by the transmit beam pattern gains at these $Q$ potential target points, defined as follows:
\begin{align}
    &B(\mathbf{h}[n],\mathbf{t}_q)=\mathbb{E}\{|\mathbf{a}^H(\mathbf{h}[n],\mathbf{z}[n],\mathbf{t}_q)\mathbf{x}[n]|^2\} \notag \\
    &\qquad= \mathbf{a}^H(\mathbf{h}[n],z[n],\mathbf{t}_q)\left(\sum_{k=1}^K\mathbf{w}_k[n]\mathbf{w}_k^H[n]+\pmb{\mathcal{R}}_s[n]\right)\notag\\
    &\qquad\quad\times\mathbf{a}(\mathbf{h}[n],z[n],\mathbf{t}_q),
\end{align}
where $\mathbf{a}(\mathbf{h}[n],z[n],\mathbf{t}_q)$ is the steering vector towards $q$th potential target as defined in (\ref{steering vector}). 

Furthermore, for SAR imaging, the SNR received at time slot $n$ can be written as~\cite{curlander1991synthetic,kim2009antenna,lahmeri2022trajectory}
\begin{align}\label{SNR}
    \mathrm{SNR}[n] =& \frac{G_tG_r\lambda^3\sigma_0c\tau_p\mathrm{PRF}\sin(\alpha)^2}{256\pi^3z[n]^3\kappa T_oNFB_rL_{tot}V}\notag\\
    &\times\left(\sum_{k=1}^K\|\mathbf{w}_k[n]\|^2 + \mathrm{tr}(\pmb{\mathcal{R}}_s[n])\right),
\end{align}
where $\kappa$ is the Boltzmann's constant, $c$ is the speed of light, $G_t$ and $G_r$ are the radar antenna gains for transmission and reception, respectively, $\sigma_0$ is the backscattering coefficient, $\tau_p$ is the radar pulse duration, $B_r$ is the radar bandwidth, $\mathrm{PRF}$ is the radar pulse repetition frequency, $NF$ is the system noise figure, $L_{tot}$ is the combined losses, and $T_o$ is the equivalent noise temperature. \begin{remark}
Different from communication systems or conventional radar modes, SAR exploits ground echoes, including those from multiple targets and even background clutter, as the very foundation of image formation rather than treating them as interference. By coherently integrating the backscattered signals from all illuminated scatterers, SAR reconstructs an image that reflects the spatial variations of their radar cross sections (RCS)~\cite{vachon2005digital}. Consequently, signals regarded as clutter in traditional radar contexts become beneficial for SAR imaging. For this reason, our analysis here focuses exclusively on the SNR.    
\end{remark}

\subsection{Energy storage}
To guarantee the long-duration flight operation of HAPS, the energy storage of the surplus power is required for future use. At time slot $n$, combining (\ref{average transmit power}) and (\ref{P_SCF}), the total energy required by the HAPS is given by \cite{javed2023interdisciplinary} 
\begin{align}
    E[n] =& (P_{\mathrm{ave}}[n]+P_{\mathrm{SCF}}[n])\Delta_t \notag\\
    \leq& \left(P_{\mathrm{max}} + \frac{V_{\mathrm{max}}^3SC_{D_0}\rho_h[n]}{2\cos(\vartheta)^2f_pf_e}\right)\Delta_t.
\end{align} 
where $V_{\mathrm{max}}=\sqrt{V_{xy\mathrm{max}}^2+V_{z\mathrm{max}}^2}$. Assume that the total available energy of the HAPS at the initial time slot is $e_{\mathrm{start}}$. To ensure continuous operation, the remaining energy at each subsequent time slot must remain nonnegative. Accordingly, the energy should satisfy the following constraint, i.e, \cite{marriott2020trajectory}
\begin{align}
     \sum_{n=1}^NE[n] \leq e_{\mathrm{start}}. 
\end{align}

\section{Problem Formulation}
\label{sec_formulation}
Assuming the ground user positions are known, we consider two deployment scenarios. In the quasi-stationary case, the HAPS remains fixed at a single optimized location, whereas in the dynamic case, it can move freely across different locations throughout all time slots. Corresponding to these deployment strategies, two optimization problems are formulated to maximize the communication metric for multiple ground users, while simultaneously satisfying the SAR imaging requirements.

\subsection{Quasi-stationary HAPS Strategy}
\begin{figure}[t]
\centering
\includegraphics[width=0.8\linewidth]{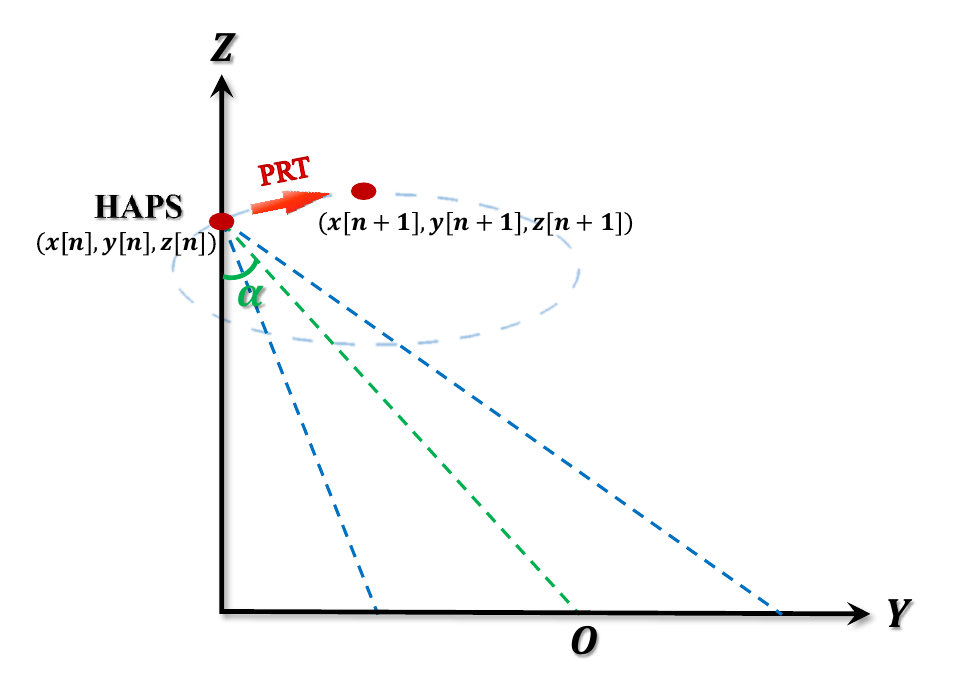}
\caption{Stop-and-go model of SAR operation.}
\label{stop_and_go}
\end{figure}

In SAR operations, the HAPS typically adopts a stop-and-go model, as illustrated in Fig.\ref{stop_and_go}. Specifically, the HAPS equipped with SAR moves to a designated position to transmit a radar pulse, pauses to receive the echo signal, and then proceeds to the next position after a pulse repetition time (PRT) interval\cite{moreira2013tutorial,liu2023integrated}. This process is repeated at each location, with the HAPS transmitting a pulse and waiting to receive the corresponding echo. Given this operational model, we adopt a quasi-stationary deployment strategy in which the HAPS remains fixed at an optimized location. For simplicity, we omit the time index $n$ and assume $\mathbf{h}[n] = \mathbf{h}$ and $z[n] = z$ for all time slots. To determine the optimal HAPS placement and beamforming configuration, we aim to maximize the weighted sum-rate throughput $\sum_{k=1}^K \beta_k r_k$ by jointly optimizing the transmit beamforming vectors $\mathbf{w}_k$ for all users, the SAR imaging covariance matrix $\pmb{\mathcal{R}}_s$, and the HAPS location variables $\mathbf{h}$ and $z$, subject to constraints on SAR imaging performance and total transmit power. Here, the weight $\beta_k > 0$ reflects the priority of $k$th user in the optimization, with larger values assigning greater importance to that user in the sum-rate maximization. Therefore, the corresponding optimization problem for the quasi-stationary HAPS strategy can be formulated as
\begin{subequations}\label{P1}
\begin{align}
\mathcal{P}1: &\mathop{\max}\limits_{\mathbf{h},z,\pmb{\mathcal{R}}_s\succeq\pmb{0},\atop \mathbf{w}_k,\forall k}\,\sum_{k=1}^K\beta_k\log_2(1+\gamma_k(\mathbf{h},z,\{\mathbf{w}_k\},\pmb{\mathcal{R}}_s)), \notag\\
&\quad\mathrm{s.t.}\,\, \mathbf{a}^H(\mathbf{h},z,\mathbf{t}_q)\left(\sum_{k=1}^K\mathbf{w}_k\mathbf{w}_k^H+\pmb{\mathcal{R}}_s\right)\mathbf{a}(\mathbf{h},z,\mathbf{t}_q)\notag\\
&\qquad\qquad\qquad\qquad\qquad\,\,\,\geq \Gamma d^2(\mathbf{h},z,\mathbf{t}_q), \,\,\forall q,  \tag{\ref{P1}a}\\
&\qquad\,\,\,\frac{G_tG_r\lambda^3\sigma_0c\tau_p\mathrm{PRF}\sin(\alpha)^2}{256\pi^3z^3\kappa T_oNFB_rL_{tot}V_{\mathrm{max}}} \notag\\
&\qquad\quad\,\,\,\times\left(\sum_{k=1}^K\|\mathbf{w}_k\|^2 + \mathrm{tr}(\pmb{\mathcal{R}}_s)\right)\geq \mathrm{SNR}_{\mathrm{min}},  \tag{\ref{P1}b} \\
&\qquad\,\,\,\sum_{k=1}^K\|\mathbf{w}_k\|^2 + \mathrm{tr}(\pmb{\mathcal{R}}_s) \leq P_{\mathrm{max}}, \tag{\ref{P1}c}
\end{align}
\end{subequations}
where $\Gamma$ is a predefined threshold. Here, (\ref{P1}a) ensures that the sensing beampattern gain at each target location meets or exceeds a threshold scaled by the square of the HAPS’s distance $d(\mathbf{h},z,\mathbf{t}_q)$. (\ref{P1}b) guarantees the minimum SNR required for SAR imaging, while (\ref{P1}c) enforces the total transmission power limit. The solution to problem $\mathcal{P}1$ will be detailed in Section~\ref{sec_quasi}.

\subsection{Dynamic HAPS Strategy}
In this scenario, we consider that a dynamic HAPS equipped with circular SAR follows a full circular trajectory over the entire imaging duration. We aim to maximize the average weighted sum-rate throughput, defined as $\frac{1}{N}\sum_{n=1}^N\sum_{k=1}^K\beta_kR_k[n]$, by jointly optimizing the transmit beamforming vectors $\mathbf{w}_k[n]$ for all users, the SAR imaging covariance matrix $\pmb{\mathcal{R}}_s[n]$, and the HAPS trajectory including $\mathbf{h}[n]$ and $z[n]$, subject to constraints on SAR imaging performance, transmit power, flight dynamics, and energy consumption across the time slots. The corresponding optimization problem for the dynamic HAPS strategy can be formulated as 
\begin{subequations}\label{P2}
\begin{align}
\mathcal{P}2:&\mathop{\max}\limits_{\mathbf{w}_k[n],\pmb{\mathcal{R}}_s[n]\succeq\pmb{0},\atop\mathbf{h}[n], z[n],\forall k,n}\,\frac{1}{N}\sum_{n=1}^N\sum_{k=1}^K\beta_kR_k[n], \notag\\
\mathrm{s.t.}&\,\, \mathbf{a}^H(\mathbf{h}[n],z[n],\mathbf{t}_q)\left(\sum_{k=1}^K\mathbf{w}_k[n]\mathbf{w}_k^H[n]+\pmb{\mathcal{R}}_s[n]\right)\notag\\
&\quad\,\times\mathbf{a}(\mathbf{h}[n],z[n],\mathbf{t}_q)\geq \Gamma d^2(\mathbf{h}[n],z[n],\mathbf{t}_q), \,\,\forall q,  \tag{\ref{P2}a}\\
&\,\frac{G_tG_r\lambda^3\sigma_0c\tau_p\mathrm{PRF}\sin(\alpha)^2}{256\pi^3H_{\mathrm{max}}^3\kappa T_oNFB_rL_{tot}V_{\mathrm{max}}} \notag\\
&\quad\,\times\left(\sum_{k=1}^K\|\mathbf{w}_k[n]\|^2 + \mathrm{tr}(\pmb{\mathcal{R}}_s[n])\right)\geq \mathrm{SNR}_{\mathrm{min}},   \tag{\ref{P2}b} \\
&\,\sum_{k=1}^K\|\mathbf{w}_k[n]\|^2 + \mathrm{tr}(\pmb{\mathcal{R}}_s[n]) \leq P_{\mathrm{max}}, \tag{\ref{P2}c} \\
&\, \sum_{n=1}^NE[n] \leq e_{\mathrm{start}}, \tag{\ref{P2}d}\\
&\, \mathbf{h}[0] = \mathbf{h}[n], z[0]=z[N], H_{\mathrm{min}}\leq z[n]\leq H_{\mathrm{max}},\tag{\ref{P2}e}\\
&\, \|\mathbf{h}[n+1]-\mathbf{h}[n]\| \leq V_{xy\mathrm{max}}\Delta_t, \tag{\ref{P2}f}\\
&\, |z[n+1]-z[n]| \leq V_{z\mathrm{max}}\Delta_t, \tag{\ref{P2}g}
\end{align}
\end{subequations}
where (\ref{P2}a) ensures that the sensing beampattern gain meets or exceeds a threshold that is proportional to the square of the HAPS's distance $d(\mathbf{h}[n],z[n],\mathbf{t}_q)$. (\ref{P2}b) guarantees that the minimum SNR required for SAR imaging is satisfied, while (\ref{P2}c) enforces the transmit power limitation. (\ref{P2}d) ensures adequate energy storage for sustained long-duration flight. Meanwhile, (\ref{P2}e), (\ref{P2}f), and (\ref{P2}g) impose feasibility and motion constraints on the HAPS trajectory. The detailed solution to problem $\mathcal{P}2$ will be presented in Section~\ref{sec_mobile}.

\section{Proposed Method for Quasi-stationary HAPS Strategy}
\label{sec_quasi}
This section presents an efficient algorithm to obtain a suboptimal solution to problem $\mathcal{P}1$. The proposed approach first applies the SCA method to jointly optimize the transmit beamforming vectors $\mathbf{w}_k$ for all users and the SAR imaging covariance matrix $\pmb{\mathcal{R}}_s$, assuming a fixed HAPS location. Subsequently, a 3D search is performed over the HAPS position variables $\mathbf{h}$ and $z$ to determine the optimal placement that maximizes the weighted sum-rate throughput.

By fixing the HAPS location $\mathbf{h}$ and altitude $z$, we formulate the following subproblem $\mathcal{P}3$, i.e.,
\begin{align}\label{P3}
\mathcal{P}3: \mathop{\max}\limits_{\mathbf{w}_k,\forall k,\atop,\pmb{\mathcal{R}}_s\succeq\pmb{0}}&\,\sum_{k=1}^K\beta_k\log_2(1+\gamma_k(\mathbf{h},z,\{\mathbf{w}_k\},\pmb{\mathcal{R}}_s)), \notag\\
\mathrm{s.t.}&\,\,\, \text{(\ref{P1}a), (\ref{P1}b), and (\ref{P1}c).} \notag
\end{align}   
To deal with the non-convex objective function in problem $\mathcal{P}3$, let $\pmb{\mathcal{W}}_k=\mathbf{w}_k\mathbf{w}_k^H$ with $\pmb{\mathcal{W}}_k\succeq\pmb{0}$ and $\mathrm{rank}(\pmb{\mathcal{W}}_k)=1$, then $\mathcal{P}3$ can be rewritten as 
\begin{subequations}\label{P4}
\begin{align}
\mathcal{P}4: &\mathop{\max}\limits_{\pmb{\mathcal{W}}_k\succeq\pmb{0},\forall k,\atop\pmb{\mathcal{R}}_s\succeq\pmb{0}}\,\sum_{k=1}^K\beta_k\hat{R}_k(\{\pmb{\mathcal{W}}_k\},\pmb{\mathcal{R}}_s),\notag\\
&\quad\,\,\,\mathrm{s.t.}\,\,\, \mathbf{a}^H(\mathbf{h},z,\mathbf{t}_q)\left(\sum_{k=1}^K\pmb{\mathcal{W}}_k+\pmb{\mathcal{R}}_s\right)\mathbf{a}(\mathbf{h},z,\mathbf{t}_q)\notag\\
&\qquad\qquad\qquad\qquad\qquad\geq \Gamma d^2(\mathbf{h},z,\mathbf{t}_q), \,\,\forall q,  \tag{\ref{P4}a}\\
&\qquad\quad\frac{G_tG_r\lambda^3\sigma_0c\tau_p\mathrm{PRF}\sin(\alpha)^2}{256\pi^3z^3\kappa T_oNFB_rL_{tot}V} \notag\\
&\qquad\quad\,\,\,\times\left(\sum_{k=1}^K\mathrm{tr}(\pmb{\mathcal{W}}_k) + \mathrm{tr}(\pmb{\mathcal{R}}_s)\right)\geq \mathrm{SNR}_{\mathrm{min}},  \tag{\ref{P4}b} \\
&\qquad\quad\sum_{k=1}^K\mathrm{tr}(\pmb{\mathcal{W}}_k) + \mathrm{tr}(\pmb{\mathcal{R}}_s) \leq P_{\mathrm{max}}, \tag{\ref{P4}c} \\
&\qquad\quad\mathrm{rank}(\pmb{\mathcal{W}}_k) \leq 1, \forall k,\tag{\ref{P4}d}
\end{align}
\end{subequations}
where $\hat{R}_k(\{\pmb{\mathcal{W}}_k\},\pmb{\mathcal{R}}_s)$ is expressed as (\ref{hat_r_k}). 

\begin{figure*}
\begin{align}\label{hat_r_k}
\hat{R}_k(\{\pmb{\mathcal{W}}_k\},\pmb{\mathcal{R}}_s)
    =& \log_2\left(\sum\limits_{k=1}^K\mathrm{tr}\left(\mathbf{g}_k(\mathbf{h},z,\mathbf{m}_k)\mathbf{g}_k^H(\mathbf{h},z,\mathbf{m}_k)\pmb{\mathcal{W}}_k\right)+\mathrm{tr}\left(\mathbf{g}_k(\mathbf{h},z,\mathbf{m}_k)\mathbf{g}_k^H(\mathbf{h},z,\mathbf{m}_k)\pmb{\mathcal{R}}_s\right) + \sigma_k^2 \right) \notag\\
    & -\log_2\left(\sum\limits_{p=1, p\neq k}^K\mathrm{tr}\left(\mathbf{g}_k(\mathbf{h},z,\mathbf{m}_k)\mathbf{g}_k^H(\mathbf{h},z,\mathbf{m}_k)\pmb{\mathcal{W}}_p\right)+\mathrm{tr}\left(\mathbf{g}_k(\mathbf{h},z,\mathbf{m}_k)\mathbf{g}_k^H(\mathbf{h},z,\mathbf{m}_k)\pmb{\mathcal{R}}_s\right) + \sigma_k^2 \right).
\end{align}   
\hrulefill
\end{figure*}

Because of the non-concave objective function and the rank constraints in (\ref{P4}a), $\mathcal{P}4$ is still a non-convex problem. Hence, we employ the SCA method~\cite{dinh2010local,zeng2017energy} to approximate the non-concave objective function iteratively as a concave one. At each iteration $o \geq 1$, we can obtain $\hat{R}_k(\{\pmb{\mathcal{W}}_k\},\pmb{\mathcal{R}}_s) \geq \bar{R}_k^{(o)}(\{\pmb{\mathcal{W}}_k\},\pmb{\mathcal{R}}_s)$, where $\bar{R}_k^{(o)}(\{\pmb{\mathcal{W}}_k\},\pmb{\mathcal{R}}_s)$ can be expressed as (\ref{R_k_iteration}) with $\bar{\mathbf{B}}_k^{(o)}$ defined as (\ref{B_k}), and  
\begin{figure*}
\begin{align}\label{R_k_iteration}
    \bar{R}_k^{(o)}(\{\pmb{\mathcal{W}}_k\},\pmb{\mathcal{R}}_s) =& \log_2\left(\sum\limits_{k=1}^K\mathrm{tr}\left(\mathbf{g}_k(\mathbf{h},z,\mathbf{m}_k)\mathbf{g}_k^H(\mathbf{h},z,\mathbf{m}_k)\pmb{\mathcal{W}}_k\right)+\mathrm{tr}\left(\mathbf{g}_k(\mathbf{h},z,\mathbf{m}_k)\mathbf{g}_k^H(\mathbf{h},z,\mathbf{m}_k)\pmb{\mathcal{R}}_s\right) + \sigma_k^2 \right) \notag\\
    &-\left(\bar{a}_k^{(o)}+\sum_{p=1,p\neq k}^K\mathrm{tr}\Big(\bar{\mathbf{B}}_k^{(o)}(\pmb{\mathcal{W}}_p-\pmb{\mathcal{W}}_p^{(o)}\Big)+\mathrm{tr}\Big(\bar{\mathbf{B}}_k^{(o)}(\pmb{\mathcal{R}}_s-\pmb{\mathcal{R}}_s^{(o)})\Big)\right), \\
\bar{\mathbf{B}}_k^{(o)} =& \frac{\log_2(e)\mathbf{g}_k(\mathbf{h},z,\mathbf{m}_k)\mathbf{g}_k^H(\mathbf{h},z,\mathbf{m}_k)}{\sum\limits_{p=1,p\neq k}^K\mathrm{tr}\left(\mathbf{g}_k(\mathbf{h},z,\mathbf{m}_k)\mathbf{g}_k^H(\mathbf{h},z,\mathbf{m}_k)\pmb{\mathcal{W}}_p^{(o)}\right)+\mathrm{tr}\left(\mathbf{g}_k(\mathbf{h},z,\mathbf{m}_k)\mathbf{g}_k^H(\mathbf{h},z,\mathbf{m}_k)\pmb{\mathcal{R}}_s^{(o)}\right)+\sigma_k^2}. \label{B_k}
\end{align}   
\hrulefill
\end{figure*}

\begin{align}
    \bar{a}_k^{(l)} =& \log_2\left(\sum\limits_{p=1\atop p\neq k}^K\mathrm{tr}\left(\mathbf{g}_k(\mathbf{h},z,\mathbf{m}_k)\mathbf{g}_k^H(\mathbf{h},z,\mathbf{m}_k)\pmb{\mathcal{W}}_p^{(l)}\right)\right.\notag\\
    &\left.+\mathrm{tr}\left(\mathbf{g}_k(\mathbf{h},z,\mathbf{m}_k)\mathbf{g}_k^H(\mathbf{h},z,\mathbf{m}_k)\pmb{\mathcal{R}}_s^{(l)}\right) + \sigma_k^2 \right). 
\end{align}
Furthermore, inspired by the SDR technique~\cite{li2024experimental,li2025target} and the approach in~\cite{lyu2022joint}, the rank constraints in (\ref{P4}d) can be relaxed as a convex SDP problem (SDR4). By iteratively solving (SDR4) using standard convex optimization tools such as CVX~\cite{grant2016cvx}, a sequence of solutions $\pmb{\mathcal{W}}_k^{(o)}$ and $\pmb{\mathcal{R}}_s^{(o)}$ are obtained, resulting in a monotonically non-decreasing objective value for problem $\mathcal{P}4$, which guarantees the convergence of the proposed SCA-and-SDR-based algorithm for solving $\mathcal{P}3$.

In summary, to enable efficient optimization of HAPS placement under the quasi-stationary strategy, the 3D deployment space is discretized into a finite set of grid points. Specifically, the horizontal coordinates $\mathbf{h} = [x, y]$ are sampled at discrete positions indexed by $n = 1, \ldots, N$, while the height $z$ is uniformly discretized over the interval $[H_{\mathrm{min}}, H_{\mathrm{max}}]$, resulting in multiple vertical layers. At each grid point, the proposed SCA-and-SDR-based algorithm is employed to maximize the weighted sum-rate throughput. The HAPS position yielding the highest objective value is selected as the optimal fixed deployment location. The overall solution for solving $\mathcal{P}1$ is summarized in Algorithm~\ref{alg_static}. Then, we analyze the computational complexity of Algorithm 1. Let $G$ denote the total number of discrete 3D grid points, and $O_{\mathrm{max}}$ represent the maximum number of iterations. At each grid point, the optimization variables include $K$ transmit beamforming matrices $\pmb{\mathcal{W}}_k$ and one SAR imaging covariance matrix $\pmb{\mathcal{R}}_s$, each of size $M\times M$. Thus, the total number of scalar decision variables is on the order of $(K+1)M^2$. According to the complexity results of interior-point methods \cite{polik2010interior,li2025scalable}, the overall computational complexity of Algorithm~1 scales on the order of $\mathcal{O}(GO_{\mathrm{max}}((K+1)M^2)^{3.5}\log(1/\epsilon))$.

\begin{algorithm}
\caption{Overall Algorithm for Solving Problem $\mathcal{P}1$}
\label{alg_static}
    \begin{algorithmic}[1]
        \STATE Discretize the 3D deployment space into a finite set of grid points. 
        \FOR{each grid point}
        \STATE Initialize $\pmb{\mathcal{W}}_k^{(0)}$, $\forall k$, and $\pmb{\mathcal{R}}_s^{(0)}$. Let $o=0$. 
        \REPEAT
        \STATE Solve problem (SDR4) for given $\pmb{\mathcal{W}}_k^{(o)}$, $\forall k$, and $\pmb{\mathcal{R}}_s^{(o)}$, and denote the optimal solution as $\pmb{\mathcal{W}}_k^{(o+1)}$, $\forall k$, and $\pmb{\mathcal{R}}_s^{(o+1)}$. 
        \STATE Update $o=o+1$. 
        \UNTIL the fractional increase of the objective value is below a threshold $\epsilon>0$.
        \ENDFOR
        \STATE Select the grid point that maximizes the sum-rate throughput as the optimal fixed HAPS location. 
    \end{algorithmic}
\end{algorithm}

\section{Proposed Method for Dynamic HAPS Strategy}
\label{sec_mobile}
This section presents an alternating optimization algorithm combined with SCA to address problem $\mathcal{P}2$. Specifically, the alternating optimization approach is implemented iteratively. In each iteration, the transmit communication beamforming vectors $\mathbf{w}_k[n]$ and the SAR imaging covariance matrix $\pmb{\mathcal{R}}_s[n]$ are optimized for a given HAPS trajectory $\mathbf{h}[n]$ and $z[n]$, followed by updating $\mathbf{h}[n]$ and $z[n]$ using the optimized $\mathbf{w}_k[n]$ and $\pmb{\mathcal{R}}_s[n]$ for $\forall k,n$. To simplify the analysis, we consider only the LOS component and neglect the NLOS paths. Accordingly, the channel can be expressed as $\mathbf{g}(\mathbf{h}[n],z[n],\mathbf{m}_k)=\ell^{-\frac{1}{2}}\mathbf{a}(\mathbf{h}[n],z[n],\mathbf{m}_k)$.

\subsection{Transmit Beamforming Optimization}
Given HAPS trajectory $\mathbf{h}[n]$ and $z[n]$, $\forall n$, we aim to optimize the communication beamforming vectors $\mathbf{w}_k[n]$ and the SAR imaging covariance matrix $\pmb{\mathcal{R}}_s$ by formulating the following problem:
\begin{align}\label{P5}
\mathcal{P}5: &\mathop{\max}\limits_{\mathbf{w}_k[n],\forall k,\atop\pmb{\mathcal{R}}_s[n]\succeq\pmb{0}}\,\frac{1}{N}\sum_{n=1}^N\sum_{k=1}^K\beta_kR_k[n], \notag\\
&\quad\,\,\mathrm{s.t.}\quad (\ref{P2}a), (\ref{P2}b), \mathrm{and}\,(\ref{P2}c).
\end{align}
Since the HAPS locations $\mathbf{h}[n]$ and $z[n]$ are fixed, the beamforming vectors $\mathbf{w}_k[n]$ and SAR imaging covariance matrix $\pmb{\mathcal{R}}_s[n]$ for $n=1,\ldots, N$ are mutually independent and can therefore be decoupled. As a result, problem $\mathcal{P}5$ can be decomposed into $N$ subproblems, each corresponding to a specific time slot $n$, formulated as follows:
\begin{align}\label{P6.n}
\mathcal{P}6.n:&\mathop{\max}\limits_{\mathbf{w}_k[n],\forall k,\atop\pmb{\mathcal{R}}_s[n]\succeq\pmb{0}}\,\sum_{k=1}^K\beta_kR_k[n], \notag\\
&\quad\,\,\mathrm{s.t.}\quad (\ref{P2}a), (\ref{P2}b), \mathrm{and}\,(\ref{P2}c).
\end{align}
It is observed that each subproblem $\mathcal{P}5.n$ has the similar structure as problem $\mathcal{P}1$, with the HAPS placement $\mathbf{h}[n]$ and $z[n]$ replacing $\mathbf{h}$ and $z$ in $\mathcal{P}1$. Therefore, the SCA-SDR hybrid method described in Section~\ref{sec_quasi} can be executed $N$ times, with each time slot solving one subproblem $\mathcal{P}6.n$. For brevity, the detailed steps are omitted.

\subsection{HAPS Trajectory Optimization}
Given the transmit communication beamforming vectors $\mathbf{w}_k[n]$ and the SAR imaging covariance matrix $\pmb{\mathcal{R}}_s[n]$, we aim at optimizing the HAPS trajectory $\mathbf{h}[n]$ and $z[n]$, which leads to the formulation of the optimization problem $\mathcal{P}7$, i.e.,  
\begin{align}
\mathcal{P}7:&\mathop{\max}\limits_{\mathbf{h}[n],z[n],\forall n}\,\frac{1}{N}\sum_{n=1}^N\sum_{k=1}^K\beta_kR_k[n], \notag\\
&\quad\,\,\,\,\mathrm{s.t.}\,\,\, \mathbf{a}^H(\mathbf{h}[n],z[n],\mathbf{t}_q)\mathbf{H}[n]\mathbf{a}(\mathbf{h}[n],z[n],\mathbf{t}_q)\notag\\
&\qquad\quad\,\, \geq \Gamma(z[n]^2+\|\mathbf{h}[n]-\mathbf{t}_q\|^2), \,\,\forall q, \notag \\
&\qquad\quad\,\,(\ref{P2}d),\,\,(\ref{P2}e),\,\, (\ref{P2}f),\,\,\mathrm{and}\,\,(\ref{P2}g),
\end{align}
where $\mathbf{H}[n]=\sum_{k=1}^K\mathbf{w}_k[n]\mathbf{w}_k^H[n]+\pmb{\mathcal{R}}_s[n]$. However, $\mathcal{P}7$ is inherently non-convex, as it involves a non-concave objective function and non-convex constraints. To address this challenge, we propose a trust-region-based SCA algorithm.

Firstly, let $\pmb{\mathcal{W}}_p[n]=\mathbf{w}_p[n]\mathbf{w}_p^H[n]$ and $\mathbf{A}(\mathbf{h}[n],z[n],\mathbf{m}_k)=\mathbf{a}(\mathbf{h}[n],z[n],\mathbf{m}_k)\mathbf{a}^H(\mathbf{h}[n],z[n],\mathbf{m}_k)$, we can re-express the $R_k[n]$ in $\mathcal{P}6$ as 
\begin{align}\label{P6_objective}
    &\hat{R}_k[n] = \log_2\left(\sum_{p=1}^Kf\left(\pmb{\mathcal{W}}_p[n],d(\mathbf{h}[n],z[n],\mathbf{m}_k)\right)\right. \notag\\
    &\,+\left.g\left(\pmb{\mathcal{R}}_s[n],d(\mathbf{h}[n],z[n],\mathbf{m}_k)\right)+\frac{\sigma_k^2}{\rho_0}d^2(\mathbf{h}[n],z[n],\mathbf{m}_k)\right) \notag\\
    &\,-\log_2\left(\sum_{p=1\atop p\neq k}^Kf\left(\pmb{\mathcal{W}}_p[n],d(\mathbf{h}[n],z[n],\mathbf{m}_k)\right)\right. \notag\\
    &\,+\left.g\left(\pmb{\mathcal{R}}_s[n],d(\mathbf{h}[n],z[n],\mathbf{m}_k)\right)+\frac{\sigma_k^2}{\rho_0}d^2(\mathbf{h}[n],z[n],\mathbf{m}_k)\right),
\end{align}
where
\begin{align}
    &f\left(\pmb{\mathcal{W}}_p[n],d(\mathbf{h}[n],z[n],\mathbf{m}_k)\right) = \mathrm{tr}(\mathbf{A}(\mathbf{h}[n],z[n],\mathbf{m}_k)\pmb{\mathcal{W}}_p[n]) \notag\\
    &= \sum_{i=1}^M\sum_{j=1}^M[\pmb{\mathcal{W}}_p[n]]_{i,j}e^{\frac{j\pi(j-i)z[n]}{d(\mathbf{h}[n],z[n],\mathbf{m}_k)}} = \mathrm{tr}(\pmb{\mathcal{W}}_p[n]) \notag \\
    &+ 2\sum_{i=1}^M\sum_{j=i+1}^M|[\pmb{\mathcal{W}}_p[n]]_{i,j}|\cos\left(\theta_{i,j}^{W_p}[n] + \frac{\pi(j-i)z[n]}{d(\mathbf{h}[n],z[n],\mathbf{m}_k)}\right),\\
    &g\left(\pmb{\mathcal{R}}_s[n],d(\mathbf{h}[n],z[n],\mathbf{m}_k)\right) = \mathrm{tr}(\pmb{\mathcal{R}}_s[n]\mathbf{A}(\mathbf{h}[n],z[n],\mathbf{m}_k)) \notag\\
    &= \sum_{i=1}^M\sum_{j=1}^M[\pmb{\mathcal{R}}_s[n]]_{i,j}e^{\frac{j\pi(j-i)z[n]}{d(\mathbf{h}[n],z[n],\mathbf{m}_k)}} = \mathrm{tr}(\pmb{\mathcal{R}}_s[n]) \notag \\
    &+ 2\sum_{i=1}^M\sum_{j=i+1}^M|[\pmb{\mathcal{R}}_s[n]]_{i,j}|\cos\left(\theta_{i,j}^{R_s}[n] + \frac{\pi (j-i)z[n]}{d(\mathbf{h}[n],z[n],\mathbf{m}_k)}\right),
\end{align}
with $\theta_{i,j}^{W_p}[n]$ and $\theta_{i,j}^{R_s}[n]$ being the phases, and $|[\pmb{\mathcal{W}}_p[n]]_{i,j}|$ and $|[\pmb{\mathcal{R}}_s[n]]_{i,j}|$ being the magnitudes of the $(i,j)$th entries of $\pmb{\mathcal{W}}_p$ and $\pmb{\mathcal{R}}_s$, respectively. Similarly, the non-convex constraints in (\ref{P2}a) can be rewritten as 
\begin{align}\label{P6_constraint}
    &\mathrm{tr}(\mathbf{H}[n]) + 2\sum_{i=1}^M\sum_{j=i+1}^M|[\mathbf{H}[n]]_{i,j}| \notag\\
    &\quad\times\cos\left(\theta_{i,j}^{H}[n] + \frac{\pi(j-i)z[n]}{d(\mathbf{h}[n],z[n],\mathbf{t}_q)}\right) \geq \Gamma d^2(\mathbf{h}[n],z[n],\mathbf{t}_q), 
\end{align}
where $\theta_{i,j}^{H}[n]$ and $|[\mathbf{H}[n]]_{i,j}|$ are the phase and magnitude of the $(i,j)$th entries of $\mathbf{H}[n]$, respectively.

Next, we use the first-order Taylor expansion to approximate the non-concave objective function, i.e.,
\begin{align}\label{Taylor_objective}
    \hat{R}_k[n] \approx \bar{R}_k[n] =& c_k^{(l)}[n] + \mathbf{v}_k^{(l)H}[n](\mathbf{h}[n]-\mathbf{h}^{(l)}[n]) \notag\\
    &+ b_k^{(l)}[n](z[n]-z^{(l)}[n]), 
\end{align}
where $c_k^{(l)}[n] = \log_2(\eta_k[n]) -\log_2(\mu_k[n])$, $\mathbf{v}_k^{(l)}[n]$ and $b_k^{(l)}[n]$ are expressed as (\ref{v_k}) and (\ref{b_k}), respectively. Here,
\begin{figure*}
\begin{align}
    &\mathbf{v}_k^{(l)}[n] = \frac{\log_2(e)}{\eta_k[n]}\left(\sum_{p=1}^K\pmb{\gamma}_v\left(\pmb{\mathcal{W}}_p[n],d(\mathbf{h}^{(l)}[n],z^{(l)}[n],\mathbf{m}_k)\right)+\pmb{\omega}_v\left(\pmb{\mathcal{R}}_s[n],d(\mathbf{h}^{(l)}[n],z^{(l)}[n],\mathbf{m}_k)\right)+\frac{2\sigma_k^2(\mathbf{h}^{(l)}[n]-\mathbf{m}_k)}{\rho_0}\right) \notag\\
    &\quad- \frac{\log_2(e)}{\mu_k[n]}\left(\sum_{p=1\atop p\neq k}^K\pmb{\gamma}_v\left(\pmb{\mathcal{W}}_p[n],d(\mathbf{h}^{(l)}[n],z^{(l)}[n],\mathbf{m}_k)\right)+\pmb{\omega}_v\left({\mathcal{R}}_s[n],d(\mathbf{h}^{(l)}[n],z^{(l)}[n],\mathbf{m}_k)\right)+\frac{2\sigma_k^2(\mathbf{h}^{(l)}[n]-\mathbf{m}_k)}{\rho_0}\right), \label{v_k} \\
    &b_k^{(l)}[n] = \frac{\log_2(e)}{\eta_k[n]}\left(\sum_{p=1}^K\gamma_b\left(\pmb{\mathcal{W}}_p[n],d(\mathbf{h}^{(l)}[n],z^{(l)}[n],\mathbf{m}_k)\right)+\omega_b\left(\pmb{\mathcal{R}}_s[n],d(\mathbf{h}^{(l)}[n],z^{(l)}[n],\mathbf{m}_k)\right)+\frac{2\sigma_k^2z^{(l)}[n]}{\rho_0}\right) \notag\\
    &\qquad\quad\,\,\,\,- \frac{\log_2(e)}{\mu_k[n]}\left(\sum_{p=1\atop p\neq k}^K\gamma_b\left(\pmb{\mathcal{W}}_p[n],d(\mathbf{h}^{(l)}[n],z^{(l)}[n],\mathbf{m}_k)\right)+\omega_b\left(\pmb{\mathcal{R}}_s[n],d(\mathbf{h}^{(l)}[n],z^{(l)}[n],\mathbf{m}_k)\right)+\frac{2\sigma_k^2z^{(l)}[n]}{\rho_0}\right). \label{b_k}
\end{align}
\hrulefill
\end{figure*}
 
\begin{align}
    &\eta_k[n] = \sum_{p=1}^Kf(\pmb{\mathcal{W}}_p[n],d(\mathbf{h}^{(l)}[n],z^{(l)}[n],\mathbf{m}_k)) \notag \\
    &\qquad\quad\,+ g(\pmb{\mathcal{R}}_s[n],d(\mathbf{h}^{(l)}[n],z^{(l)}[n],\mathbf{m}_k)) \notag\\
    &\qquad\quad\,+ \frac{\sigma_k^2}{\rho_0}d^2(\mathbf{h}^{(l)}[n],z^{(l)}[n],\mathbf{m}_k), \\
    &\mu_k[n] = \sum_{p=1\atop p\neq 
 k}^Kf(\pmb{\mathcal{W}}_p[n],d(\mathbf{h}^{(l)}[n],z^{(l)}[n],\mathbf{m}_k)) \notag \\
    &\qquad\quad\,+ g(\pmb{\mathcal{R}}_s[n],d(\mathbf{h}^{(l)}[n],z^{(l)}[n],\mathbf{m}_k)) \notag\\
    &\qquad\quad\,+ \frac{\sigma_k^2}{\rho_0}d^2(\mathbf{h}^{(l)}[n],z^{(l)}[n],\mathbf{m}_k), \\
&\pmb{\gamma}_v\left(\pmb{\mathcal{W}}_p[n],d(\mathbf{h}^{(l)}[n],z^{(l)}[n],\mathbf{m}_k)\right), \notag\\
    &\quad= \sum_{i=1}^M\sum_{j=i+1}^M2\pi|[\pmb{\mathcal{W}}_p[n]]_{i,j}|\sin\left(\theta_{i,j}^{W_p}[n] \right. \notag\\
    &\qquad\left.+\frac{\pi (j-i)z^{(l)}[n]}{d(\mathbf{h}^{(l)}[n],z^{(l)}[n],\mathbf{m}_k)}\right)\frac{(j-i)z^{(l)}[n](\mathbf{h}^{(l)}[n]-\mathbf{m}_k)}{d^3(\mathbf{h}^{(l)}[n],z^{(l)}[n],\mathbf{m}_k)}, \\
    &\pmb{\omega}_v\left(\pmb{\mathcal{R}}_s[n],d(\mathbf{h}^{(l)}[n],z^{(l)}[n],\mathbf{m}_k)\right) \notag\\
    &\quad= \sum_{i=1}^M\sum_{j=i+1}^M2\pi|[\pmb{\mathcal{R}}_s[n]]_{i,j}|\sin\left(\theta_{i,j}^{R_s}[n]\right. \notag\\
    &\qquad\left.+\frac{\pi (j-i)z^{(l)}[n]}{d(\mathbf{h}^{(l)}[n],z^{(l)}[n],\mathbf{m}_k)}\right)\frac{(j-i)z^{(l)}[n](\mathbf{h}^{(l)}[n]-\mathbf{m}_k)}{d^3(\mathbf{h}^{(l)}[n],z^{(l)}[n],\mathbf{m}_k)}, 
\end{align}
    
\begin{align}       &\gamma_b\left(\pmb{\mathcal{W}}_p[n],d(\mathbf{h}^{(l)}[n],z^{(l)}[n],\mathbf{m}_k)\right) \notag\\
    &\quad= \sum_{i=1}^M\sum_{j=i+1}^M2\pi|[\pmb{\mathcal{W}}_p[n]]_{i,j}|\sin\left(\theta_{i,j}^{W_p}[n] \right. \notag\\
    &\qquad\left.+\frac{\pi(j-i)z^{(l)}[n]}{d(\mathbf{h}^{(l)}[n],z^{(l)}[n],\mathbf{m}_k)}\right)\frac{(j-i)\|\mathbf{h}^{(l)}[n]-\mathbf{m}_k\|^2}{d^3(\mathbf{h}^{(l)}[n],z^{(l)}[n],\mathbf{m}_k)}, \\
    &\omega_b\left(\pmb{\mathcal{R}}_s[n],d(\mathbf{h}^{(l)}[n],z^{(l)}[n],\mathbf{m}_k)\right) \notag\\
    &\quad= \sum_{i=1}^M\sum_{j=i+1}^M2\pi|[\pmb{\mathcal{R}}_s[n]]_{i,j}|\sin\left(\theta_{i,j}^{R_s}[n]\right. \notag\\
    &\qquad\left.+\frac{\pi(j-i)z^{(l)}[n]}{d(\mathbf{h}^{(l)}[n],z^{(l)}[n],\mathbf{m}_k)}\right)\frac{(j-i)\|\mathbf{h}^{(l)}[n]-\mathbf{m}_k\|^2}{d^3(\mathbf{h}^{(l)}[n],z^{(l)}[n],\mathbf{m}_k)}. 
\end{align}
Similarly, we use the first-order Taylor expansion to approximate the non-convex constraints in ($\mathcal{P}7$), i.e., 
\begin{align}\label{Taylor_constraint}
    &\Xi_j^{(l)}[n] + \pmb{\nu}_{j}^{(l)H}[n](\mathbf{h}[n]-\mathbf{h}^{(l)}[n]) + \zeta_{j}^{(l)}[n](z[n]-z^{(l)}[n])\notag\\
    &\geq \Gamma(z^2[n]+\|\mathbf{h}[n]-\mathbf{t}_q\|^2), 
\end{align}
where 
\begin{align}
    \Xi_j^{(l)}[n] &= \mathrm{tr}(\mathbf{H}[n]) + 2\sum_{i=1}^M\sum_{j=i+1}^M|[\mathbf{H}[n]]_{i,j}| \notag\\
    &\times\cos\left(\theta_{i,j}^{H}[n] + \frac{\pi (j-i)z^{(l)}[n]}{d(\mathbf{h}^{(l)}[n],z^{(l)}[n],\mathbf{t}_q)}\right), \\
    \pmb{\nu}_j^{(l)}[n] &= \sum_{i=1}^M\sum_{j=i+1}^M2\pi|[\mathbf{H}[n]]_{i,j}|\sin\left(\theta_{i,j}^{H}[n]  \right. \notag\\
    &\left.+ \frac{\pi(j-i)z^{(l)}[n]}{d(\mathbf{h}^{(l)}[n],z^{(l)}[n],\mathbf{t}_q)}\right)\frac{(j-i)z^{(l)}[n](\mathbf{h}^{(l)}[n]-\mathbf{t}_q)}{d^3(\mathbf{h}^{(l)}[n],z^{(l)}[n],\mathbf{t}_q)}, \\
    \zeta_j^{(l)}[n] &= \sum_{i=1}^M\sum_{j=i+1}^M2\pi|[\mathbf{H}[n]]_{i,j}|\sin\left(\theta_{i,j}^{H}[n] \right. \notag\\
    &\left.+\frac{\pi(j-i)z^{(l)}[n]}{d(\mathbf{h}^{(l)}[n],z^{(l)}[n],\mathbf{t}_q)}\right)\frac{(j-i)\|\mathbf{h}^{(l)}[n]-\mathbf{t}_q\|^2}{d^3(\mathbf{h}^{(l)}[n],z^{(l)}[n],\mathbf{t}_q)}.
\end{align}
At this stage, we have approximated the non-concave objective function (\ref{P6_objective}) and the non-convex constraint (\ref{P6_constraint}) using their respective forms in (\ref{Taylor_objective}) and (\ref{Taylor_constraint}). To ensure the accuracy of these approximations, we impose a set of trust region constraints as
\begin{align}\label{trust region}
    \|\mathbf{h}^{(l)}[n]-\mathbf{h}^{(l-1)}[n]\| \leq \varphi^{(l)},\,\,|z^{(l)}[n]-z^{(l-1)}[n]| \leq \varphi^{(l)}, 
\end{align}
where $\varphi^{(l)}$ is the radius of the trust region. 

Finally, by combining (\ref{Taylor_objective}), (\ref{Taylor_constraint}), and (\ref{trust region}), we obtain the convex approximation of problem $\mathcal{P}7$ in the $l$th iteration, denoted as problem $\mathcal{P}8.l$, i.e.,
\begin{align}
    &\mathcal{P}8.l: \max_{\mathbf{h}[n],z[n],\forall n} \frac{1}{N}\sum_{n=1}^N\sum_{k=1}^K\beta_k\bar{R}_k^{(l)}[n],\notag\\
    &\qquad\quad\mathrm{s.t.}\,\,\, (\ref{Taylor_constraint}), (\ref{trust region}), (\ref{P2}d), (\ref{P2}e), (\ref{P2}f), \mathrm{and}\,\, (\ref{P2}g),
\end{align}
which can be efficiently solved to optimality using CVX. Note that theoretically, each iteration is guaranteed to converge if the trust region radius $\varphi^{(l)}$ is sufficiently small~\cite{conn2000trust}. In practical implementation, however, if solving $\mathcal{P}8.l$ in the $l$th iteration does not yield a lower objective value for $\mathcal{P}7$ compared to the previous iteration, we reduce the trust region radius to $\varphi^{(l)}=\varphi^{(l)}/2$ and resolve $\mathcal{P}8.l$. The iteration process terminates when $\varphi^{(l)}$ falls below a predefined threshold $\epsilon$. 

\begin{algorithm}[t]
\caption{Overall Algorithm for Solving Problem $\mathcal{P}_2$}
\label{alg_mobile}
\begin{algorithmic}[1]
    \STATE Initialize $\pmb{\mathcal{W}}_k^{(0)}[n]$, $\pmb{\mathcal{R}}_s^{(0)}[n]$, $\mathbf{h}^{(0)}[n]$ and $\mathbf{z}^{(0)}[n]$, $\forall k,n$. Let $o=0$. 
    \REPEAT 
    \STATE Solve the problem $\mathcal{P}5.n$ for given $\pmb{\mathcal{W}}_k^{(o)}[n]$, and $\pmb{\mathcal{R}}_s^{(o)}[n]$, $\mathbf{h}^{(o)}[n]$ and $z^{(o)}[n]$, and denote the optimal solution as $\pmb{\mathcal{W}}_k^{(o+1)}[n]$, and $\pmb{\mathcal{R}}_s^{(o+1)}[n]$, $\forall k,n$.
    \STATE Let $l=0$, $\mathbf{h}^{(l)}[n]=\mathbf{h}^{(o)}[n]$, and $z^{(l)}=z^{(o)}$.  
    \REPEAT
    \STATE Solve the problem $\mathcal{P}8.l$ for given $\pmb{\mathcal{W}}_k^{(o+1)}[n]$, and $\pmb{\mathcal{R}}_s^{(o+1)}[n]$ to obtain $\mathbf{h}^{(l+1)}[n]$ and $z^{(l+1)}[n]$, $\forall k,n$.
    \IF{the objective value of the problem $\mathcal{P}7$ increases}     
    \STATE Update $l=l+1$.
    \ELSE 
        \STATE Execute $\varphi^{(l)}=\varphi^{(l)}/2$. 
    \ENDIF
    \UNTIL{$\varphi^{(l)}<\epsilon$}
    \STATE Update $o=o+1$. 
    \UNTIL{the fractional increase of the objective value is below a threshold $\epsilon$}.
\end{algorithmic}
\end{algorithm}

\subsection{Summary}
The overall SCA-alternating optimization algorithm for solving problem $\mathcal{P}2$ is outlined in Algorithm~\ref{alg_mobile}. This algorithm iteratively integrates the transmit beamforming optimization from Section~\ref{sec_mobile}-A and the HAPS trajectory optimization from Section~\ref{sec_mobile}-B. The outer iteration is indexed by $o$. In each outer iteration, the transmit beamforming subproblem $\mathcal{P}5.n$ is solved using the SCA-and-SDR-based method described in Section~\ref{sec_quasi}, which ensures convergence. Subsequently, the HAPS trajectory subproblem $\mathcal{P}7$ is addressed using a trust-region-based SCA algorithm, which also guarantees convergence. Through this alternating procedure, the objective value of $\mathcal{P}2$ monotonically increases and remains upper-bounded by a finite constant at each outer iteration. Consequently, the convergence of Algorithm~\ref{alg_mobile} is guaranteed. We also note that, due to the inherent non-convexity of the problem and the use of SCA, the proposed algorithm converges to a stationary point, i.e., a locally optimal solution. Although global optimality is difficult to guarantee mathematically, the effectiveness of the obtained solution is validated through extensive comparisons with multiple benchmark schemes in our simulation results.

Finally, we analyze the computational complexity of Algorithm~2. Let $O_{\mathrm{max}}$ and $I_{\mathrm{max}}$ denote the maximum number of iterations for the outer and inner layer, respectively. In the beamforming stage, each time slot involves $(K+1)M^2$ scalar decision variables, and solving the $N$ subproblems requires $\mathcal{O}(O_{\mathrm{max}}(N((K+1)M^2)^{3.5}\log(1/\epsilon))$ operations per outer iteration. In the trajectory stage, there are $3N$ optimization variables, leading to a complexity cost of $\mathcal{O}(I_{\mathrm{max}}(3N)^{3.5}\log(1/\epsilon))$ per outer iteration. Therefore, the overall computational complexity of Algorithm~2 is $\mathcal{O}(O_{\mathrm{max}}(N((K+1)M^2)^{3.5}+I_{\mathrm{max}}(3N)^{3.5})\log(1/\epsilon))$.

\section{Simulation results}    
\label{sec_sim}
In this section, numerical examples are provided to demonstrate the effectiveness and advantages of the proposed algorithm. We consider an ISAC-enabled HAPS system serving $K=8$ ground users and sensing $Q=8$ targets, both randomly and uniformly distributed within a two-dimensional (2D) imaging area of $4\times4$ $\mathrm{km}^2$, with each occupying half of the area, respectively. The HAPS operates with a maximum horizontal flight speed of $V_{xy\mathrm{max}} = 40$ m/s and a maximum vertical flight speed of $V_{z\mathrm{max}} = 30$ m/s. Its flight altitude is constrained between $H_{\mathrm{min}} = 20$ km and $H_{\mathrm{max}} = 30$ km. The HAPS is equipped with $M=12$ antennas. Additionally, the channel power gain at the reference distance $d_0=1$ is set as $\rho_0=30$ dB, while the noise power gain at each user receiver is fixed at $\sigma_k^2=-60$ dBm. For simplicity and fairness, the weight coefficients are chosen as $\beta_k = 1$, $\forall k$. The Rician factor $K_u$ is set to 10. Detailed system parameters related to aerodynamics and SAR imaging are summarized in Tables \ref{parameters_HAPS} and \ref{parameters_SAR}, respectively~\cite{javed2023interdisciplinary,lahmeri2022trajectory}.

\subsection{Quasi-Stationary HAPS Strategy}
\begin{figure}[t]
\centering
\includegraphics[width=0.9\linewidth]{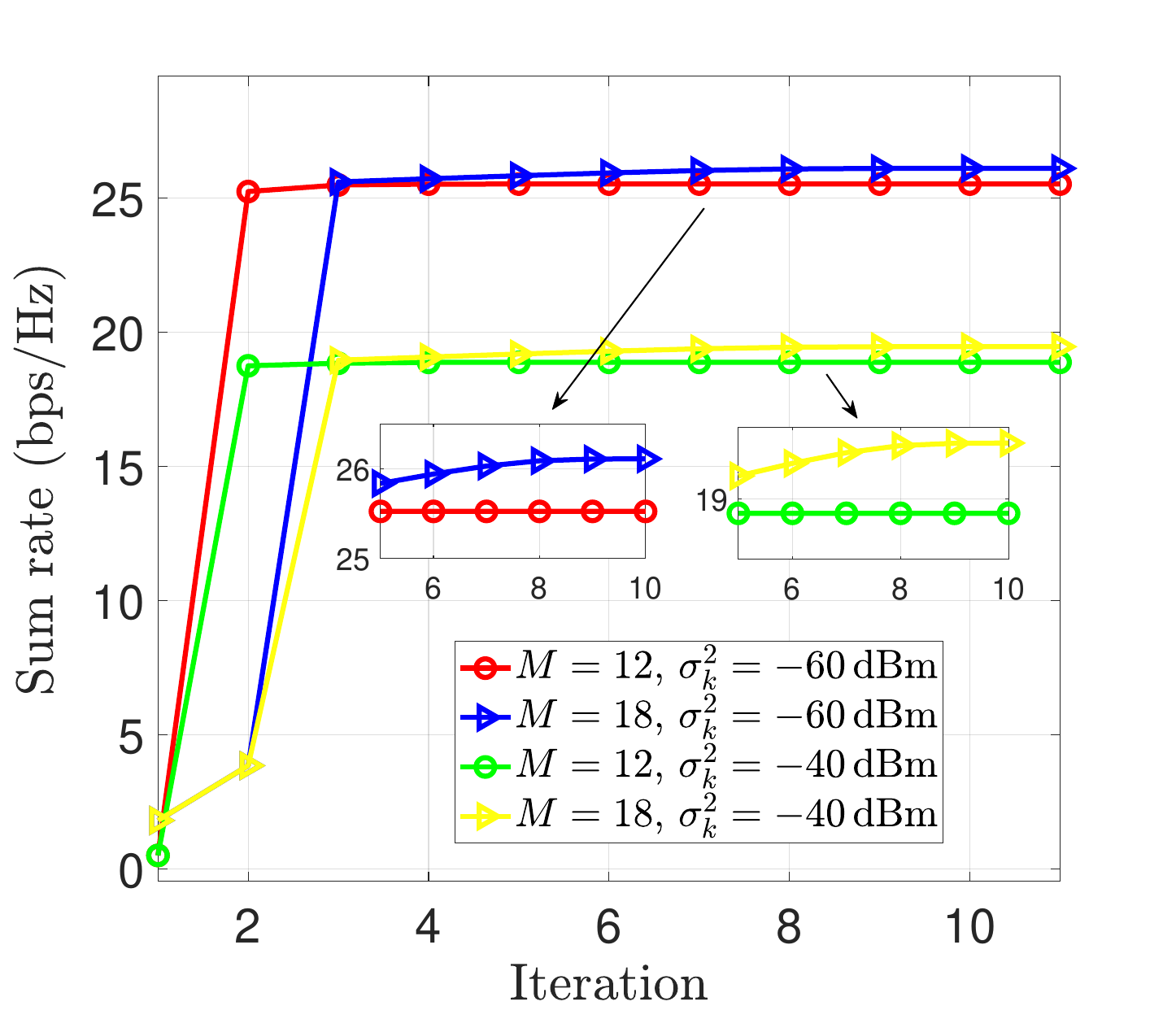}
\caption{Convergence of the proposed algorithm~\ref{alg_static}.}
\label{convergence_static}
\end{figure}

Before presenting the performance comparison, we first examine the convergence behavior of the proposed Algorithm~\ref{alg_static}. In this evaluation, the beampattern gain threshold is set to $\Gamma=-36$\,dBm, and the maximum transmit power is fixed at $P_{\mathrm{max}}=10$\,W. As illustrated in Fig.~\ref{convergence_static}, under different numbers of antennas $M$ and noise power levels $\sigma_k^2$, the sum-rate throughput achieved by the proposed algorithm increases steadily and converges rapidly within a few iterations, thereby demonstrating its efficient convergence performance.

\begin{table}[t]
\centering
\caption{Simulation Parameters for Aerodynamics}
\label{parameters_HAPS}
\renewcommand\arraystretch{1.1}
\begin{tabular}{|c|c|}
\hline
Parameter & Default Value \\ \hline
$C_{D_0}$ & 0.015 \\ \hline
$S$ & 143\,$m^2$ \\ \hline
$f_p$, $f_e$ & 0.85, 0.90 \\ \hline
$e_o$ & 0.6385 \\ \hline
$AR_{\omega}$ & 30 \\ \hline
$F_{\omega}$ & 165\,kg \\ \hline
\end{tabular}
\end{table}

\begin{table}[t]
\centering
\caption{Simulation Parameters for SAR Imaging}
\label{parameters_SAR}
\renewcommand\arraystretch{1.1}
\begin{tabular}{|c|c|}
\hline
Parameter & Default Value \\ \hline
$\alpha$ & $45^{\circ}$ \\ \hline
$G_t$, $G_r$ & 35\,dB  \\ \hline
$\sigma_0$ & 1 \\ \hline
$\tau_p$ & 10\,$g$s \\ \hline
$\mathrm{PRF}$ & 2\,Hz \\ \hline
$\kappa$ & $1.380649\times10^{-23}$\,J/K \\ \hline
$T_0$ & 290\,K \\ \hline
$NF$ & 6\,dB\\ \hline
$B_r$ & 200\,MHz\\ \hline
$L_{tot}$ & 10\,dB\\ \hline
\end{tabular}
\end{table}

\begin{figure}[t]
\centering
\subfloat[3D view]{
\includegraphics[width=0.9\linewidth]{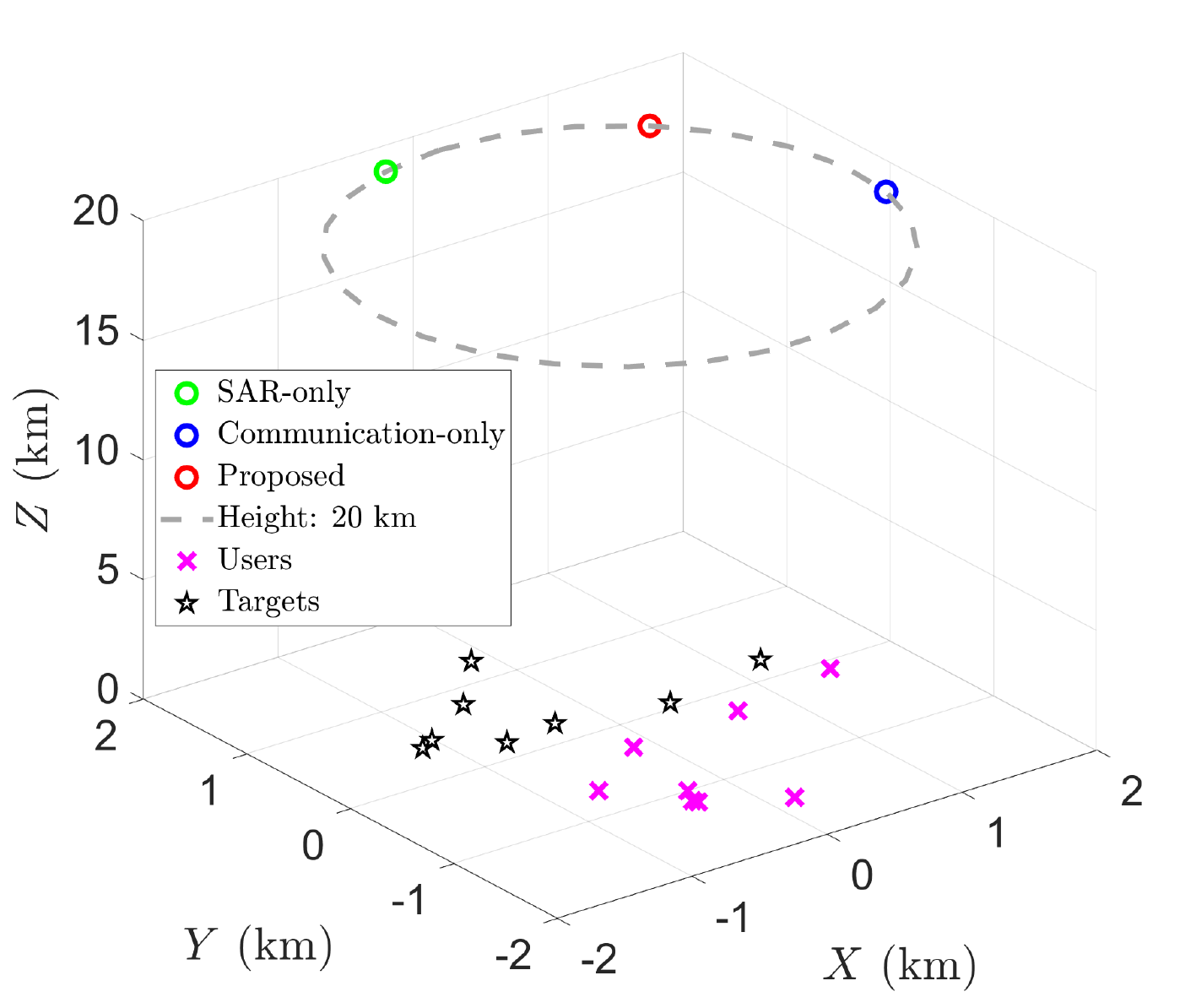}}
\\
\subfloat[Top-down 2D view]{
\includegraphics[width=0.9\linewidth]{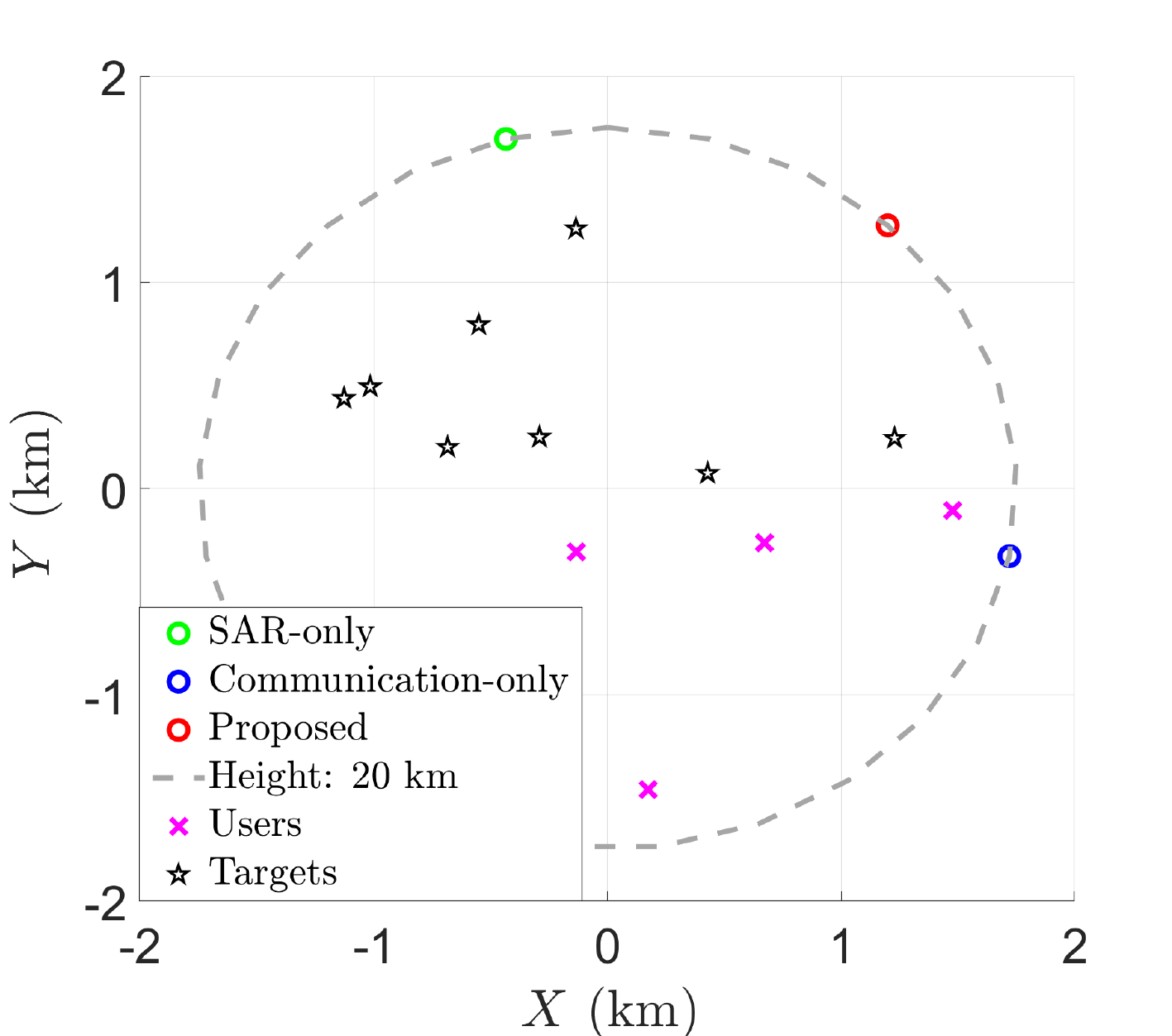}}
\caption{Obtained deployment locations of the HAPS in (a) 3D space and (b) the 2D horizontal plane..}
\label{optimal_deployment}
\end{figure}

For comparison purposes, we then employ the following benchmark.
\begin{itemize}
    \item \textbf{Communication-Only Beamforming Design}: The HAPS is dedicated solely to the communication task, corresponding to solving the following optimization problem:
\begin{align}
\mathcal{P}9: &\mathop{\max}\limits_{\mathbf{w}_k,\forall k}\,\sum_{k=1}^K\beta_k\log_2(1+\gamma_k(\mathbf{h},z,\{\mathbf{w}_k\})), \notag\\
&\,\,\,\,\mathrm{s.t.}\,\,\sum_{k=1}^K\|\mathbf{w}_k\|^2 \leq P_{\mathrm{max}}. \notag
\end{align} 
    \item \textbf{SAR-Only Beamforming Design}: The HAPS is dedicated solely to the SAR imaging task, corresponding to solving the following optimization problem:
\begin{align}
\mathcal{P}10: &\max_{\atop\pmb{\mathcal{R}}_s\succeq \pmb{0}}\,\min_{j} \frac{\mathbf{a}^H(\mathbf{h},z,\mathbf{t}_q)\pmb{\mathcal{R}}_s\mathbf{a}(\mathbf{h},z,\mathbf{t}_q)}{d^2(\mathbf{h},z,\mathbf{t}_q)}, \notag \\
&\,\,\,\,\mathrm{s.t.}\,\,\,\frac{G_tG_r\lambda^3\sigma_0c\tau_p\mathrm{PRF}\sin(\alpha)^2}{256\pi^3z^3\kappa T_oNFB_rL_{tot}V_{\mathrm{max}}} \notag\\
&\qquad\qquad\qquad\times\mathrm{tr}(\pmb{\mathcal{R}}_s) \geq \mathrm{SNR}_{\mathrm{min}}, \notag \\
&\qquad\,\,\,\, \mathrm{tr}(\pmb{\mathcal{R}}_s)\leq P_{\mathrm{max}}. \notag
\end{align}
\end{itemize}
Problems $\mathcal{P}9$ and $\mathcal{P}10$ can be solved following the same methodology as $\mathcal{P}1$, and the details are omitted for brevity. We perform a 3D search over a 2D horizontal plane using $25$ candidate points and discretize the flight altitude within the range of $[20,30]$ km in $1$ km intervals. Fig.~\ref{optimal_deployment} illustrates the resulting HAPS deployment locations for three scenarios: SAR-only scheme, communication-only scheme, and our proposed ISAC scheme. In all cases, the optimal deployment is found at the lowest altitude level. For the sensing-only design, the HAPS is positioned directly over the sensing area, whereas in the communication-only design, it is located closer to the communication users. In contrast, the proposed ISAC design places the HAPS between the communication and sensing regions, indicating that our algorithm effectively balances the trade-off between communication and SAR imaging performance. 

\begin{figure}
\centering
\includegraphics[width=0.9\linewidth]{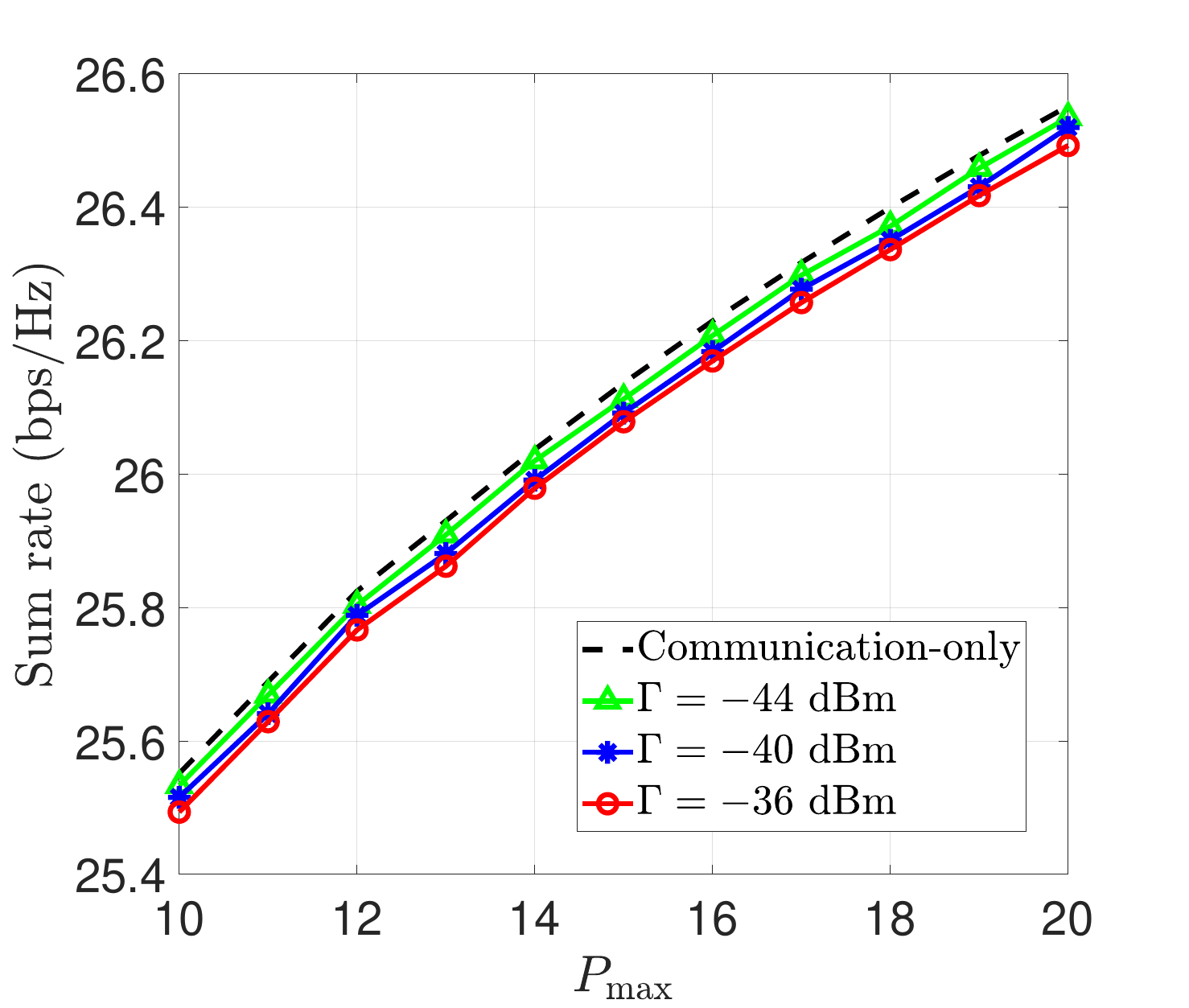}
\caption{The optimal sum-rate throughput versus versus the maximum transmit power $P_{\mathrm{max}}$ under different $\Gamma$ for quasi-stationary HAPS.}
\label{power_static}
\end{figure}

Furthermore, we evaluate the optimal sum-rate throughput of both the communication-only design and the proposed ISAC design as a function of the maximum transmit power $P_{\mathrm{max}}$ at the HAPS, under sensing beampattern gain constraints $\Gamma$ set to $-36$\,dBm, $-40$\,dBm, and $-44$\,dBm, respectively. As shown in Fig.~\ref{power_static}, the sum-rate throughput increases monotonically with $P_{\mathrm{max}}$ across all scenarios. Additionally, as $\Gamma$ increases, the impact of the sensing constraint on communication performance diminishes, and the optimal sum-rate throughput of the ISAC design gradually converges toward that of the communication-only design. It is also observed that variations in $\Gamma$ do not lead to significant fluctuations in the optimal sum-rate throughput. This is primarily because $\Gamma$ mainly influences the beamforming direction rather than the total transmit power available for communication. Moreover, due to the high-altitude and wide-coverage characteristics of the HAPS, these directional adjustments have only a limited effect on the communication link gain, ensuring that the system throughput remains relatively stable across different sensing constraint levels. Consequently, the optimal HAPS location may shift to balance sensing requirements with communication performance, thereby highlighting the inherent trade-off in ISAC system design.

\begin{figure}[t]
\centering
\includegraphics[width=0.9\linewidth]{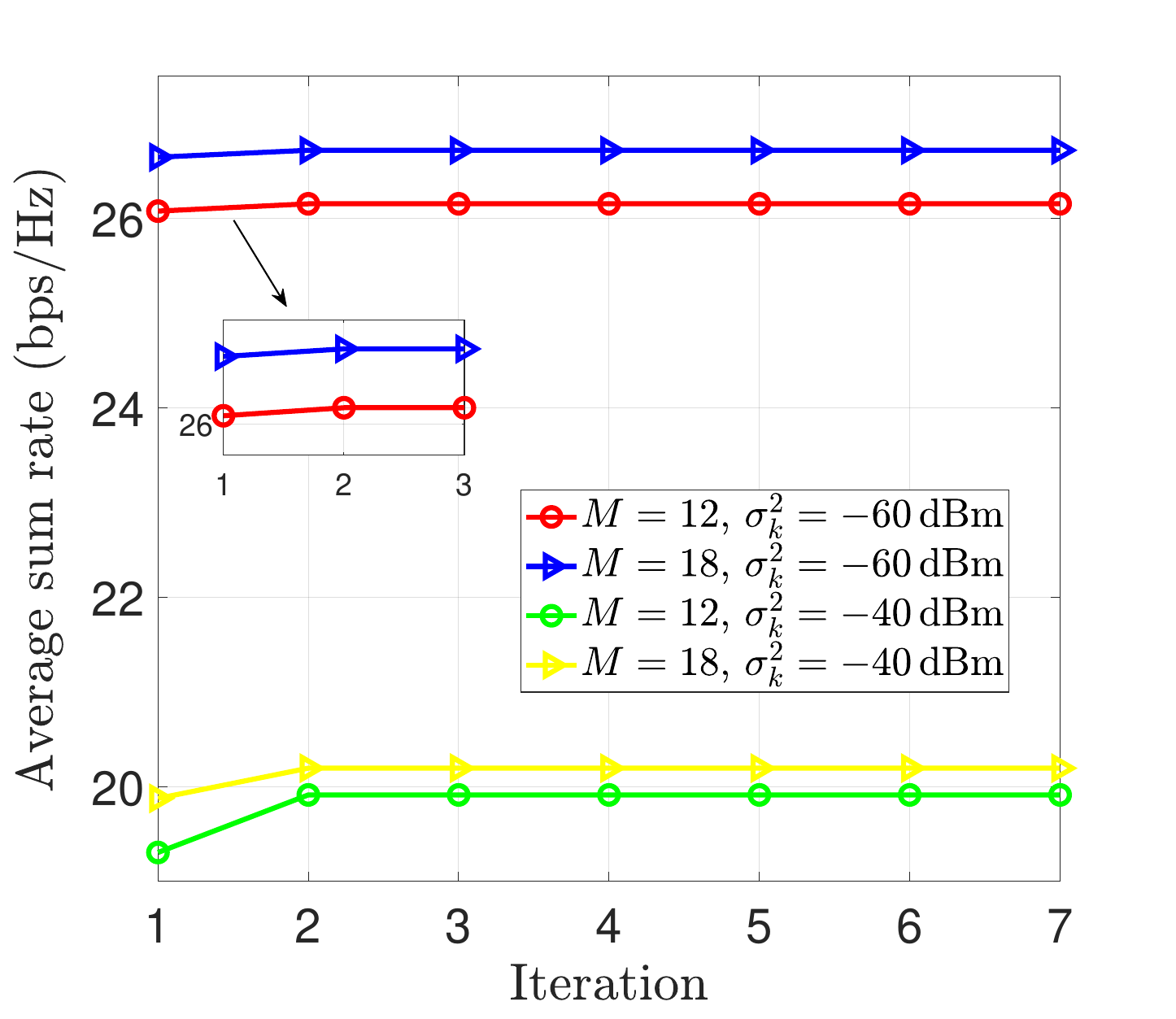}
\caption{Convergence of the proposed algorithm~\ref{alg_mobile}.}
\label{convergence_mobile}
\end{figure}

\begin{figure}[t!]
\centering
\subfloat[3D view]{
\includegraphics[width=0.9\linewidth]{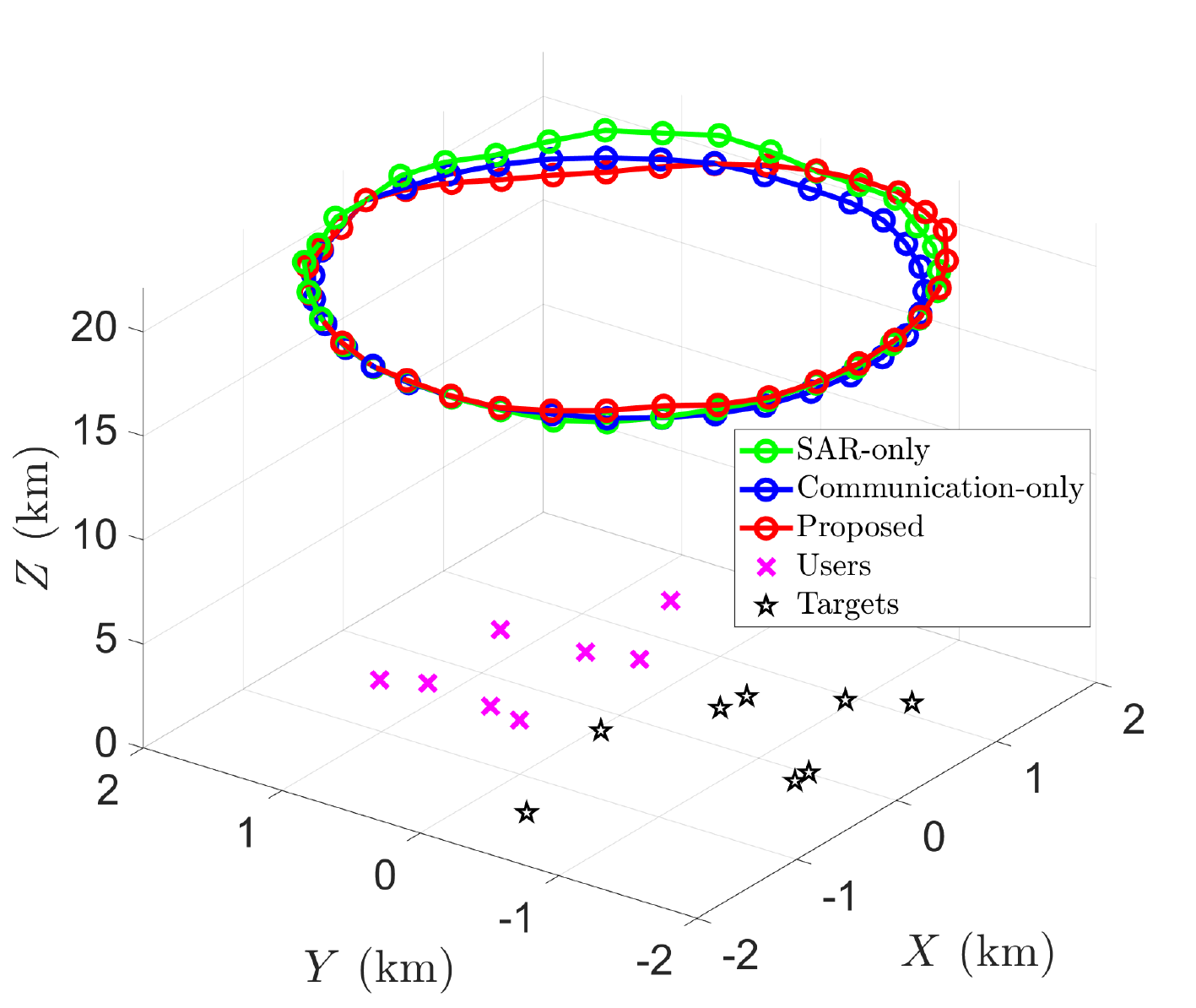}}
\\
\subfloat[Top-down 2D view]{
\includegraphics[width=0.9\linewidth]{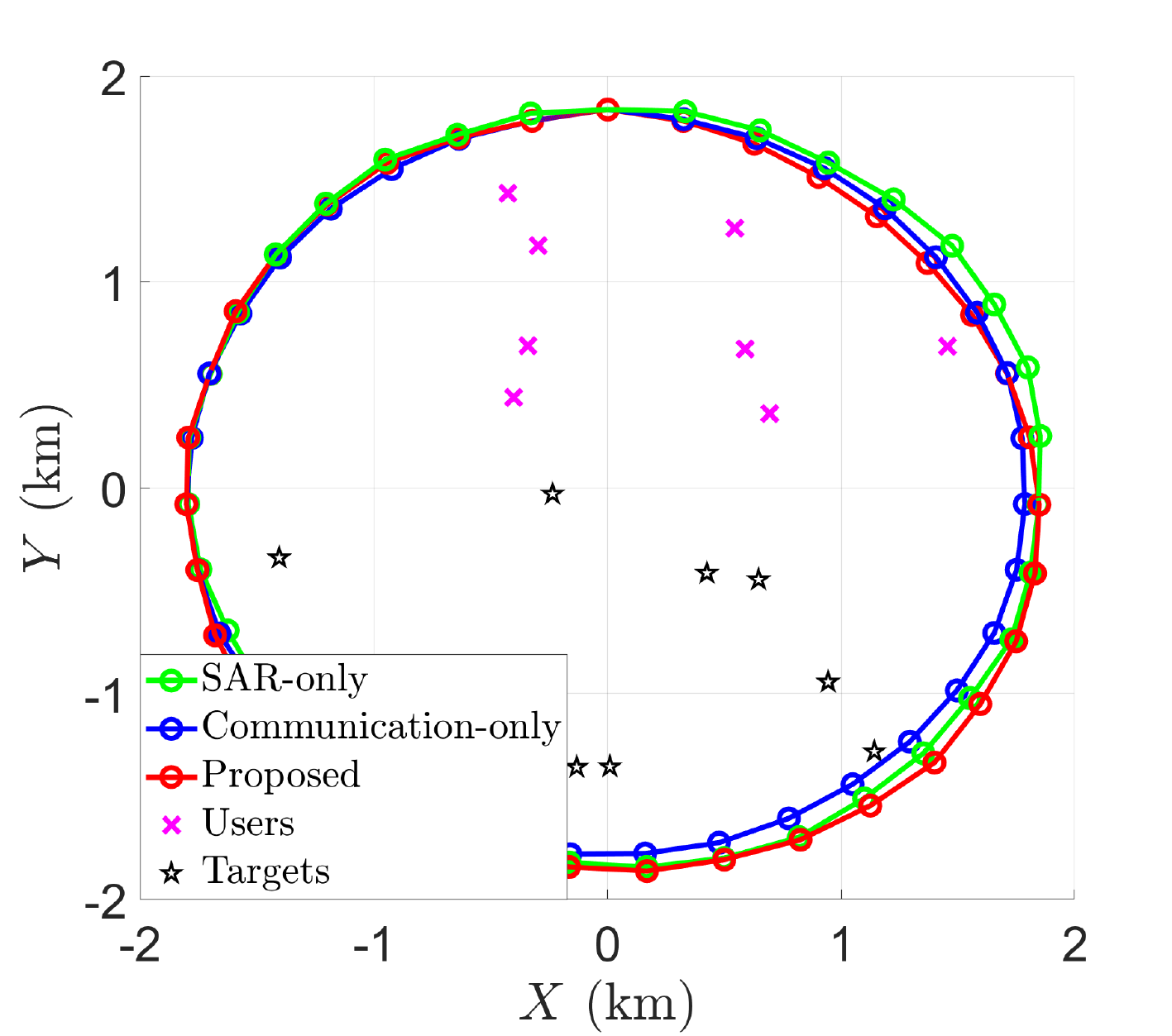}}
\caption{Obtained trajectories of the HAPS in (a) 3D space and (b) the 2D horizontal plane.}
\label{optimal_trajectory}
\end{figure}

\subsection{Dynamic HAPS Strategy}
We consider a period $T=350$ s with a sample interval of $\Delta_t=10$ s. Before comparing performance, we first examine the convergence behavior of the proposed Algorithm~\ref{alg_mobile}. The beampattern gain threshold is set to $\Gamma=-50$ dBm, and the maximum transmit power is assumed to be $P_{\mathrm{max}}=10$ W. With different numbers of antennas $M$ and noise power levels $\sigma_k^2$, Fig.~\ref{convergence_mobile} shows the average sum-rate throughput achieved by the proposed algorithm converges rapidly. This fast convergence is attributed to a sufficiently large number of inner loop iterations, which enables the outer loop to converge within just two iterations.

\begin{figure}
\centering
\includegraphics[width=0.9\linewidth]{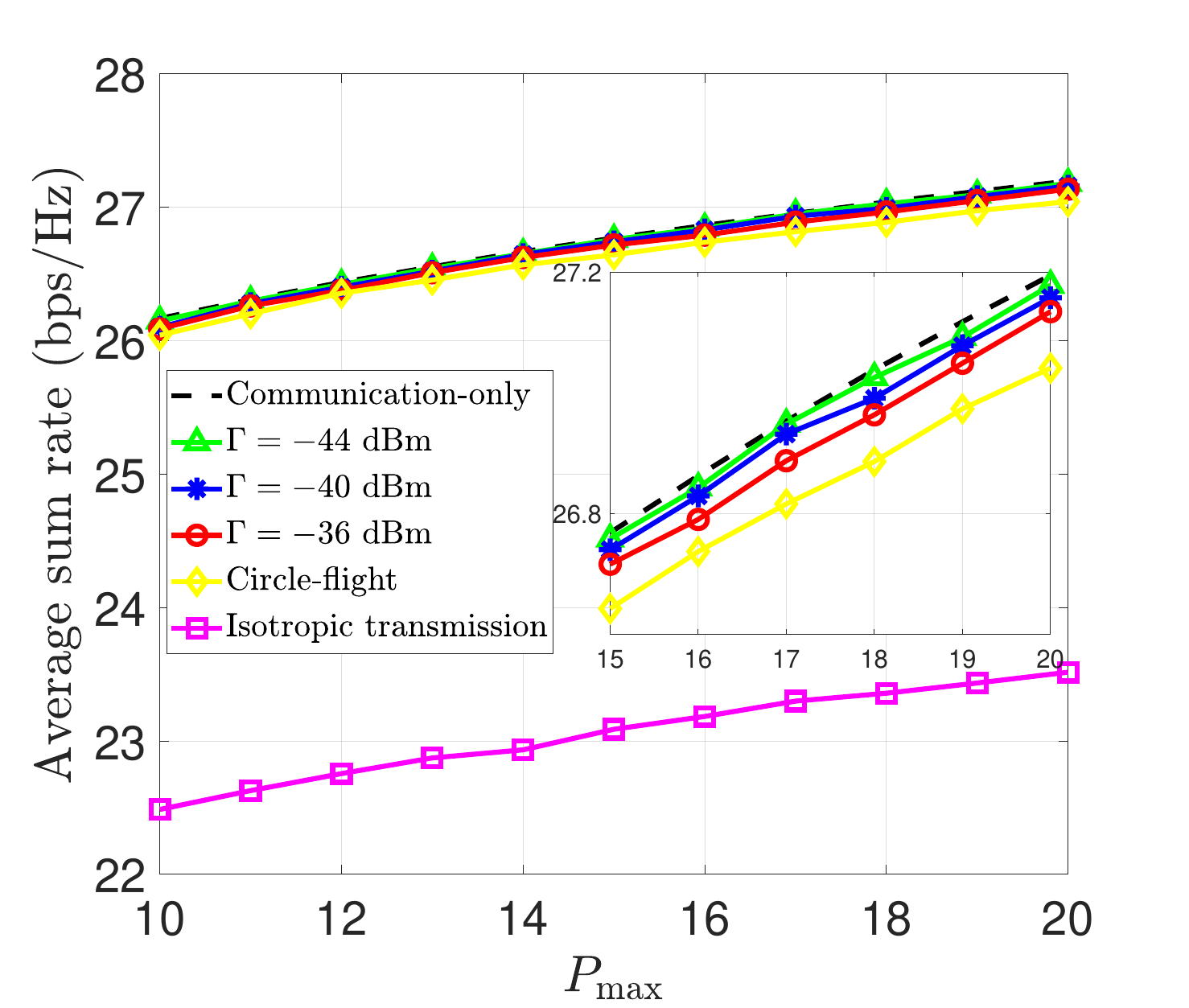}
\caption{The optimal average sum-rate throughput versus the maximum transmit power $P_{\mathrm{max}}$ under different $\Gamma$ for dynamic HAPS.}
\label{power_mobile}
\end{figure}

Then we adopt the following benchmark for comparison.
\begin{itemize}
    \item 
    \textbf{Circle-Flight Beamforming Design}: The HAPS follows a randomly generated closed 3D circular trajectory at its allowable speed. Along this path, both the horizontal position and altitude vary in a coordinated manner. Given the predetermined circular trajectory $\mathbf{h}[n]$ and $z[n]$, the transmit beamforming vectors $\mathbf{w}_k[n]$ and the SAR covariance matrices $\pmb{\mathcal{R}}_s[n]$, $\forall k,n$ are jointly optimized by solving problem $\mathcal{P}3$.
    \item \textbf{Isotropic Transmission Design}: The HAPS adopts isotropic transmission, where the beamforming matrices are defined as $\pmb{\mathcal{W}}_k[n] = \frac{P_{c,k}}{M}\mathbf{I}_M$ and $\pmb{\mathcal{R}}_s[n] = \frac{P_t}{M}\mathbf{I}_M$, $\forall k,n$, with $P_{c,k}$ and $P_t$ denoting the transmit powers allocated to the communication and dedicated SAR imaging signals, respectively. Accordingly, the total power constraint is given by $\sum_{k=1}^K P_{c,k} + P_t \leq P_{\mathrm{max}}$. By substituting $\pmb{\mathcal{W}}_k[n]$ and $\pmb{\mathcal{R}}_s[n]$ into problem~$\mathcal{P}1$, the joint design simplifies to a power allocation and trajectory optimization problem, which can be efficiently solved through iterative optimization over the transmit power and trajectory variables. 
    \item \textbf{Communication-Only Beamforming Design}: The HAPS is dedicated solely to the communication task, corresponding to solving the following optimization problem:
    \begin{subequations}
\begin{align}
\mathcal{P}11: &\mathop{\max}\limits_{\mathbf{w}_k[n],\pmb{\mathcal{R}}_s[n]\succeq\pmb{0},\atop\mathbf{h}[n], z[n],\forall k,n}\,\frac{1}{N}\sum_{n=1}^N\sum_{k=1}^K\beta_kR_k[n], \notag\\
&\qquad\,\,\mathrm{s.t.}\quad\sum_{k=1}^K\|\mathbf{w}_k[n]\|^2 \leq P_{\mathrm{max}},  \notag\\
&\qquad\qquad\,\,\,(\ref{P2}d),\,(\ref{P2}e),\,(\ref{P2}f),\,\mathrm{and}\,(\ref{P2}g). \notag
\end{align}
\end{subequations}
    \item \textbf{SAR Imaging-Only Beamforming Design}: The HAPS is dedicated solely to the SAR imaging task, which corresponds to solving the following optimization problem:
    \begin{align}
        &\mathcal{P}12: \notag\\
        &\max_{\mathbf{h}[n],z[n],\forall n\atop\pmb{\mathcal{R}}_s[n]\succeq \pmb{0}} \min_{q} \frac{\mathbf{a}^H(\mathbf{h}[n],z[n],\mathbf{t}_q)\pmb{\mathcal{R}}_s[n]\mathbf{a}(\mathbf{h}[n],z[n],\mathbf{t}_q)}{d^2(\mathbf{h}[n],z[n],\mathbf{t}_q)}, \notag \\
        &\qquad\mathrm{s.t.}\quad\frac{G_tG_r\lambda^3\sigma_0c\tau_p\mathrm{PRF}\sin(\alpha)^2}{256\pi^3H_{\mathrm{max}}^3\kappa T_oNFB_rL_{tot}V_{\mathrm{max}}}\notag\\
&\qquad\qquad\qquad\qquad\times\mathrm{tr}(\pmb{\mathcal{R}}_s[n])\geq \mathrm{SNR}_{\mathrm{min}}, \notag\\
        &\qquad\qquad\,\,\,\mathrm{tr}(\pmb{\mathcal{R}}_s[n])\leq P_{\mathrm{max}}, \notag\\, 
        &\qquad\qquad\,\,\,(\ref{P2}d),\,(\ref{P2}e),\,(\ref{P2}f),\,\mathrm{and}\,(\ref{P2}g). \notag
    \end{align}
\end{itemize}
Problems $\mathcal{P}11$ and $\mathcal{P}12$ can be solved using the same methodology as $\mathcal{P}2$, and the detailed derivations are therefore omitted for brevity. The optimized HAPS trajectories for three representative scenarios are illustrated in Fig.~\ref{optimal_trajectory}, corresponding to the SAR-only scheme, the communication-only scheme, and the proposed ISAC scheme. Furthermore, we evaluate the average sum-rate throughput achieved by the proposed ISAC design, the circle-flight design, the isotropic transmission design and the communication-only design as a function of the maximum transmit power $P_{\mathrm{max}}$ at the HAPS, under different sensing beampattern gain constraints $\Gamma$. As shown in Fig.~\ref{power_mobile}, the average sum-rate throughput increases monotonically with $P_{\mathrm{max}}$ for all schemes, where the proposed ISAC design consistently achieves higher throughput than the circle-flight and isotropic-transmission baselines across the entire transmit-power range. Moreover, as $\Gamma$ increases, the impact of the sensing constraint on communication performance diminishes, and the optimal sum-rate throughput of the proposed ISAC design gradually converges toward that of the communication-only design. We can also observe that variations in $\Gamma$ do not lead to significant fluctuations in the optimal average sum-rate throughput. This is primarily because $\Gamma$ mainly influences the beamforming direction rather than the total transmit power available for communication.

\par

\section{Conclusion}
\label{sec_con}
In this paper, we explored an ISAC-enabled HAPS system that simultaneously facilitates multi-user communication and ground target sensing over a designated SAR imaging area. To enhance communication throughput while satisfying SAR imaging requirements, we considered two HAPS deployment strategies: quasi-stationary and dynamic operation. For each strategy, we formulated a sensing-constrained sum-rate maximization problem and developed efficient algorithms to jointly optimize the 3D HAPS placement or trajectory alongside the transmit beamforming for both communication and sensing tasks. Numerical results highlight the critical role of HAPS deployment design in effectively managing the trade-off between communication performance and SAR imaging quality.

\appendices

\bibliographystyle{IEEEtran}
\bibliography{references}

@article{abbasi2024haps,
  title={{HAPS} for {6G} networks: Potential use cases, open challenges, and possible solutions},
  author={Abbasi, Omid and Yadav, Animesh and Yanikomeroglu, Halim and Dao, Ngoc-Dng and Senarath, Gamini and Zhu, Peiying},
  journal={IEEE Wireless Communications},
  year={2024},
  publisher={IEEE}
}

@article{kurt2021vision,
  title={A vision and framework for the high altitude platform station ({HAPS}) networks of the future},
  author={Kurt, Gunes Karabulut and Khoshkholgh, Mohammad G and Alfattani, Safwan and Ibrahim, Ahmed and Darwish, Tasneem SJ and Alam, Md Sahabul and Yanikomeroglu, Halim and Yongacoglu, Abbas},
  journal={IEEE Communications Surveys \& Tutorials},
  volume={23},
  number={2},
  pages={729--779},
  year={2021},
  publisher={IEEE}
}

@article{mohammed2011role,
  title={The role of high-altitude platforms ({HAPs}) in the global wireless connectivity},
  author={Mohammed, Abbas and Mehmood, Asad and Pavlidou, Fotini-Niovi and Mohorcic, Mihael},
  journal={Proceedings of the IEEE},
  volume={99},
  number={11},
  pages={1939--1953},
  year={2011},
  publisher={IEEE}
}

@article{liu2022integrated,
  title={Integrated sensing and communications: Toward dual-functional wireless networks for {6G} and beyond},
  author={Liu, Fan and Cui, Yuanhao and Masouros, Christos and Xu, Jie and Han, Tony Xiao and Eldar, Yonina C and Buzzi, Stefano},
  journal={IEEE journal on selected areas in communications},
  volume={40},
  number={6},
  pages={1728--1767},
  year={2022},
  publisher={IEEE}
}

@article{wang2014high,
  title={High altitude platform multichannel {SAR} for wide-area and staring imaging},
  author={Wang, Wen-Qin and Shao, Huaizong},
  journal={IEEE Aerospace and Electronic Systems Magazine},
  volume={29},
  number={5},
  pages={12--17},
  year={2014},
  publisher={IEEE}
}

@article{hu2022trajectory,
  title={Trajectory planning of cellular-connected {UAV} for communication-assisted radar sensing},
  author={Hu, Shuyan and Yuan, Xin and Ni, Wei and Wang, Xin},
  journal={IEEE Transactions on Communications},
  volume={70},
  number={9},
  pages={6385--6396},
  year={2022},
  publisher={IEEE}
}

@article{lyu2022joint,
  title={Joint maneuver and beamforming design for {UAV}-enabled integrated sensing and communication},
  author={Lyu, Zhonghao and Zhu, Guangxu and Xu, Jie},
  journal={IEEE Transactions on Wireless Communications},
  volume={22},
  number={4},
  pages={2424--2440},
  year={2022},
  publisher={IEEE}
}

@article{arum2020energy,
  title={Energy management of solar-powered aircraft-based high altitude platform for wireless communications},
  author={Arum, Steve Chukwuebuka and Grace, David and Mitchell, Paul Daniel and Zakaria, Muhammad Danial and Morozs, Nils},
  journal={Electronics},
  volume={9},
  number={1},
  pages={179},
  year={2020},
  publisher={MDPI}
}

@article{javed2023interdisciplinary,
  title={An Interdisciplinary Approach to Optimal Communication and Flight Operation of High-Altitude Long-Endurance Platforms},
  author={Javed, Sidrah and Alouini, Mohamed-Slim and Ding, Zhiguo},
  journal={IEEE Transactions on Aerospace and Electronic Systems},
  year={2023},
  publisher={IEEE}
}

@book{stengel2004flight,
  title={Flight dynamics},
  author={Stengel, Robert F},
  year={2004},
  publisher={Princeton, NJ, USA: Princeton Univ. Press}
}

@inproceedings{lahmeri2022trajectory,
  title={Trajectory and resource optimization for {UAV} synthetic aperture radar},
  author={Lahmeri, Mohamed-Amine and Ghanem, Walid and Knill, Christina and Schober, Robert},
  booktitle={2022 IEEE Globecom Workshops (GC Wkshps)},
  pages={897--903},
  year={2022},
  organization={IEEE}
}

@article{tan2022joint,
  title={Joint communication and {SAR} waveform design method via time-frequency spectrum shaping},
  author={Tan, Youshan and Li, Zhongyu and Yang, Jing and Yu, Xianxiang and An, Hongyang and Wu, Junjie and Yang, Jianyu},
  journal={IEEE Transactions on Geoscience and Remote Sensing},
  volume={60},
  pages={1--13},
  year={2022},
  publisher={IEEE}
}

@article{yang2022waveform,
  title={Waveform design for watermark framework based {DFRC} system with application on joint SAR imaging and communication},
  author={Yang, Jing and Tan, Youshan and Yu, Xianxiang and Cui, Guolong and Zhang, Di},
  journal={IEEE Transactions on Geoscience and Remote Sensing},
  volume={61},
  pages={1--14},
  year={2022},
  publisher={IEEE}
}

@article{wang2019first,
  title={First demonstration of joint wireless communication and high-resolution {SAR} imaging using airborne {MIMO} radar system},
  author={Wang, Jie and Liang, Xing-Dong and Chen, Long-Yong and Wang, Li-Na and Li, Kun},
  journal={IEEE Transactions on Geoscience and Remote Sensing},
  volume={57},
  number={9},
  pages={6619--6632},
  year={2019},
  publisher={IEEE}
}

@misc{gsma2021high,
  title={High Altitude Platform Systems-Towers in the Skies},
  author={GSMA, M},
  year={2021},
  publisher={GSMA}
}

@INPROCEEDINGS{marriott2020trajectory,
  author={Marriott, Jack and Tezel, Birce and Liu, Zhang and Stier-Moses, Nicolas E.},
  booktitle={2020 6th International Conference on Control, Automation and Robotics (ICCAR)}, 
  title={Trajectory Optimization of Solar-Powered High-Altitude Long Endurance Aircraft}, 
  year={2020},
  pages={473-481},
}

@inproceedings{azzahra2019noma,
  title={{NOMA} signal transmission over millimeter-wave frequency for backbone network in {HAPS} with {MIMO} antenna},
  author={Azzahra, Mirrah Aliya and others},
  booktitle={2019 IEEE 13th international conference on telecommunication systems, services, and applications (TSSA)},
  pages={186--189},
  year={2019},
  organization={IEEE}
}

@article{ji2020energy,
  title={Energy-efficient beamforming for beamspace {HAP-NOMA} systems},
  author={Ji, Pingping and Jiang, Lingge and He, Chen and Lian, Zhuxian and He, Di},
  journal={IEEE Communications Letters},
  volume={25},
  number={5},
  pages={1678--1681},
  year={2020},
  publisher={IEEE}
}

@inproceedings{hsieh2020uav,
  title={{UAV}-based multi-cell {HAPS} communication: System design and performance evaluation},
  author={Hsieh, Frank and Jardel, Fanny and Visotsky, Eugene and Vook, Frederick and Ghosh, Amitava and Picha, Bob},
  booktitle={GLOBECOM 2020-2020 IEEE Global Communications Conference},
  pages={1--6},
  year={2020},
  organization={IEEE}
}

@inproceedings{nauman2017system,
  title={System design and performance evaluation of high altitude platform: Link budget and power budget},
  author={Nauman, Ali and Maqsood, Moazam},
  booktitle={2017 19th International Conference on Advanced Communication Technology (ICACT)},
  pages={138--142},
  year={2017},
  organization={IEEE}
}

@article{hua20203d,
  title={{3D UAV} trajectory and communication design for simultaneous uplink and downlink transmission},
  author={Hua, Meng and Yang, Luxi and Wu, Qingqing and Swindlehurst, A Lee},
  journal={IEEE Transactions on Communications},
  volume={68},
  number={9},
  pages={5908--5923},
  year={2020},
  publisher={IEEE}
}

@inproceedings{dinh2010local,
  title={Local convergence of sequential convex programming for nonconvex optimization},
  author={Dinh, Quoc Tran and Diehl, Moritz},
  booktitle={Recent Advances in Optimization and its Applications in Engineering: The 14th Belgian-French-German Conference on Optimization},
  pages={93--102},
  year={2010},
  organization={Springer}
}

@article{zeng2017energy,
  title={Energy-efficient {UAV} communication with trajectory optimization},
  author={Zeng, Yong and Zhang, Rui},
  journal={IEEE Transactions on wireless communications},
  volume={16},
  number={6},
  pages={3747--3760},
  year={2017},
  publisher={IEEE}
}

@misc{grant2016cvx,
  title={{CVX}: {Matlab} software for disciplined convex programming},
  author={Grant, Michael and Boyd, Stephen},
  year={2016},
  note={[Online] Available: http://cvxr.com/cvx}
}

@article{wu2018joint,
  title={Joint trajectory and communication design for multi-{UAV} enabled wireless networks},
  author={Wu, Qingqing and Zeng, Yong and Zhang, Rui},
  journal={IEEE Transactions on Wireless Communications},
  volume={17},
  number={3},
  pages={2109--2121},
  year={2018},
  publisher={IEEE}
}

@article{kim2009antenna,
  title={Antenna mask design for {SAR} performance optimization},
  author={Kim, Se Young and Myung, Noh Hoon and Kang, Min Jeong},
  journal={IEEE Geoscience and Remote Sensing Letters},
  volume={6},
  number={3},
  pages={443--447},
  year={2009},
  publisher={IEEE}
}

@book{curlander1991synthetic,
  title={Synthetic aperture radar},
  author={Curlander, John C and McDonough, Robert N},
  volume={11},
  year={1991},
  publisher={Wiley, New York}
}

@article{zheng2024random,
  title={Random signal design for joint communication and {SAR} imaging towards low-altitude economy},
  author={Zheng, Bowen and Liu, Fan},
  journal={IEEE Wireless Communications Letters},
  year={2024},
  publisher={IEEE}
}

@book{conn2000trust,
  title={Trust region methods},
  author={Conn, Andrew R and Gould, Nicholas IM and Toint, Philippe L},
  year={2000},
  publisher={SIAM}
}

@article{yahia2022haps,
  title={{HAPS} selection for hybrid {RF/FSO} satellite networks},
  author={Yahia, Olfa Ben and Erdogan, Eylem and Kurt, Gunes Karabulut and Altunbas, Ibrahim and Yanikomeroglu, Halim},
  journal={IEEE Transactions on Aerospace and Electronic Systems},
  volume={58},
  number={4},
  pages={2855--2867},
  year={2022},
  publisher={IEEE}
}

@article{benaya2025aerial,
  title={Aerial {ISAC}: A {HAPS}-assisted integrated sensing, communications and computing framework for enhanced coverage and security},
  author={Benaya, Ahmed M and Hassan, Mohamed S and Ismail, Mahmoud H and Landolsi, Taha},
  journal={IEEE Transactions on Green Communications and Networking},
  year={2025},
  publisher={IEEE}
}

@article{kaushik2024integrated,
  title={Integrated sensing and communications for {IoT}: Synergies with key {6G} technology enablers},
  author={Kaushik, Aryan and Singh, Rohit and Li, Ming and Luo, Honghao and Dayarathna, Shalanika and Senanayake, Rajitha and An, Xueli and Stirling-Gallacher, Richard A and Shin, Wonjae and Di Renzo, Marco},
  journal={IEEE Internet of Things Magazine},
  year={2024},
  publisher={IEEE}
}

@article{moreira2013tutorial,
  title={A tutorial on synthetic aperture radar},
  author={Moreira, Alberto and Prats-Iraola, Pau and Younis, Marwan and Krieger, Gerhard and Hajnsek, Irena and Papathanassiou, Konstantinos P},
  journal={IEEE Geoscience and remote sensing magazine},
  volume={1},
  number={1},
  pages={6--43},
  year={2013},
  publisher={IEEE}
}

@inproceedings{liu2023integrated,
  title={Integrated sensing and communication for {UAV}-borne {SAR} systems},
  author={Liu, Ziyi and Zesong, Fei and Liu, Peng and Wang, Xinyi and Zheng, Zhong and Zhou, Dongkai and Yuan, Weijie},
  booktitle={2023 22nd International Symposium on Communications and Information Technologies (ISCIT)},
  pages={1--6},
  year={2023},
  organization={IEEE}
}

@article{li2011influence,
  title={The influence of target micromotion on {SAR} and {GMTI}},
  author={Li, Xiang and Deng, Bin and Qin, Yuliang and Wang, Hongqiang and Li, Yanpeng},
  journal={IEEE Transactions on Geoscience and Remote Sensing},
  volume={49},
  number={7},
  pages={2738--2751},
  year={2011},
  publisher={IEEE}
}

@article{makhoul2014performance,
  title={A performance evaluation of {SAR-GMTI} missions for maritime applications},
  author={Makhoul, Eduardo and Broquetas, Antoni and Rodon, Josep Ruiz and Zhan, Yu and Ceba, Francisco},
  journal={IEEE Transactions on Geoscience and Remote Sensing},
  volume={53},
  number={5},
  pages={2496--2509},
  year={2014},
  publisher={IEEE}
}

@article{soumekh1996reconnaissance,
  title={Reconnaissance with slant plane circular {SAR} imaging},
  author={Soumekh, Mehrdad},
  journal={IEEE transactions on image processing},
  volume={5},
  number={8},
  pages={1252--1265},
  year={1996},
  publisher={IEEE}
}

@article{shamsabadi2024enhancing,
  title={Enhancing next-generation urban connectivity: Is the integrated {HAPS}-terrestrial network a solution?},
  author={Shamsabadi, Afsoon Alidadi and Yadav, Animesh and Yanikomeroglu, Halim},
  journal={IEEE Communications Letters},
  volume={28},
  number={5},
  pages={1112--1116},
  year={2024},
  publisher={IEEE}
}

@article{abbasi2024hemispherical,
  title={Hemispherical antenna array architecture for high-altitude platform stations ({HAPS}) for uniform capacity provision},
  author={Abbasi, Omid and Yanikomeroglu, Halim and Kaddoum, Georges},
  journal={IEEE Transactions on Wireless Communications},
  year={2024},
  publisher={IEEE}
}

@incollection{polik2010interior,
  title={Interior point methods for nonlinear optimization},
  author={P{\'o}lik, Imre and Terlaky, Tam{\'a}s},
  booktitle={Nonlinear Optimization: Lectures given at the CIME Summer School held in Cetraro, Italy, July 1-7, 2007},
  pages={215--276},
  year={2010},
  publisher={Springer}
}

@article{vachon2005digital,
  title={Digital Processing of Synthetic Aperture Radar Data: Algorithms and implementation.},
  author={Vachon, Paris W},
  journal={Geomatica},
  volume={59},
  number={3},
  pages={353--355},
  year={2005}
}

@inproceedings{zhang2024joint,
  title={Joint robust secure beamforming designs for {ISAC}-enabled {LEO} satellite systems},
  author={Zhang, Xue and Wang, Ruibo and Shang, Bodong and Alouini, Mohamed-Slim},
  booktitle={ICC 2024-IEEE International Conference on Communications},
  pages={1182--1188},
  year={2024},
  organization={IEEE}
}

@article{wang2024network,
  title={Network-Level Analysis of Integrated Sensing and Communication Using Stochastic Geometry},
  author={Wang, Ruibo and Belmekki, Baha Eddine Youcef and Zhang, Xue and Alouini, Mohamed-Slim},
  journal={IEEE Internet of Things Magazine},
  volume={7},
  number={4},
  pages={84--90},
  year={2024},
  publisher={IEEE}
}

@article{zhang2025ris,
  title={{RIS}-Based {DOA} Estimation for Communication-Assisted Sensing Systems Under Hardware Impairments},
  author={Zhang, Xue and Le, Ngoc Phuc and Alouini, Mohamed-Slim},
  journal={IEEE Open Journal of Vehicular Technology},
  year={2025},
  publisher={IEEE}
}

@article{huang2025design,
  title={Design of frequency index modulated waveforms for integrated {SAR} and communication on high-altitude platforms ({HAPs})},
  author={Huang, Bang and Ahmed, Sajid and Alouini, Mohamed-Slim},
  journal={IEEE Transactions on Communications},
  year={2025},
  publisher={IEEE}
}

@article{wang2014mimo,
  title={{MIMO SAR OFDM} chirp waveform diversity design with random matrix modulation},
  author={Wang, Wen-Qin},
  journal={IEEE Transactions on Geoscience and Remote Sensing},
  volume={53},
  number={3},
  pages={1615--1625},
  year={2014},
  publisher={IEEE}
}

@article{li2025target,
  title={Target Localization With Unknown Transmit Power using Rank-One Semidefinite Programming},
  author={Li, Yingquan and Xu, Jiajie and Mukhopadhyay, Bodhibrata and Alouini, Mohamed-Slim},
  journal={IEEE Wireless Communications Letters},
  year={2025},
  volume={14},
  number={11},
  pages={3450-3454},
  publisher={IEEE}
}

@article{li2024experimental,
  title={Experimental validation of cooperative RSS-based localization with unknown transmit power, path loss exponent, and precise anchor location},
  author={Li, Yingquan and Mukhopadhyay, Bodhibrata and Xu, Jiajie and Alouini, Mohamed-Slim},
  journal={IEEE Transactions on Wireless Communications},
  year={2024},
  volume={23},
  number={11},
  pages={16482-16497},
  publisher={IEEE}
}

@article{li2025scalable,
  title={Scalable Cooperative Localization using Augmented Lagrangian Method with Experimental Validation},
  author={Li, Yingquan and Mukhopadhyay, Bodhibrata and Kammoun, Abla and Alouini, Mohamed-Slim},
  journal={IEEE Transactions on Wireless Communications},
  year={2025},
  note={Early access},
  publisher={IEEE}
}

\end{document}